\documentclass[acmsmall,authorversion]{acmart}
\usepackage{enumitem}
\usepackage{makecell}
\usepackage{multirow}

\setcopyright{acmlicensed}
\acmJournal{PACMMOD}
\acmYear{2025} \acmVolume{3} \acmNumber{6 (SIGMOD)}
\acmArticle{294} \acmMonth{12} \acmPrice{}
\acmDOI{10.1145/3769759}

\begin{document}

\title[Analysis and Evaluation of Using Microsecond-Latency Memory in SSD-Based Key-Value Stores]{Analysis and Evaluation of Using Microsecond-Latency Memory for In-Memory Indices and Caches in SSD-Based Key-Value Stores}

\newcommand\kioxia{\institution{Kioxia Corporation}\city{Yokohama}\country{Japan}}
\newcommand\fixstars{\institution{Fixstars Corporation}\city{Tokyo}\country{Japan}}

\author{Yosuke Bando}
\affiliation{%
  \kioxia
}
\email{yosuke1.bando@kioxia.com}

\author{Akinobu Mita}
\affiliation{%
  \fixstars
}
\email{mita@fixstars.com}

\author{Kazuhiro Hiwada}
\affiliation{%
  \kioxia
}
\email{kazuhiro.hiwada@kioxia.com}

\author{Shintaro Sano}
\affiliation{%
  \kioxia
}
\email{shintarou.sano@kioxia.com}

\author{Tomoya Suzuki}
\affiliation{%
  \kioxia
}
\email{tomoya.suzuki@kioxia.com}

\author{Yu Nakanishi}
\affiliation{%
  \kioxia
}
\email{yu.nakanishi@kioxia.com}

\author{Kazutaka Tomida}
\affiliation{%
  \kioxia
}
\email{kazutaka1.tomida@kioxia.com}

\author{Hirotsugu Kajihara}
\affiliation{%
  \kioxia
}
\email{hirotsugu.kajihara@kioxia.com}

\author{Akiyuki Kaneko}
\affiliation{%
  \kioxia
}
\email{akiyuki.kaneko@kioxia.com}

\author{Daisuke Taki}
\affiliation{%
  \kioxia
}
\email{daisuke.taki@kioxia.com}

\author{Yukimasa Miyamoto}
\affiliation{%
  \kioxia
}
\email{yukimasa.miyamoto@kioxia.com}

\author{Tomokazu Yoshida}
\affiliation{%
  \fixstars
}
\email{tomokazu.yoshida@fixstars.com}

\author{Tatsuo Shiozawa}
\affiliation{%
  \kioxia
}
\email{tatsuo.shiozawa@kioxia.com}

\renewcommand{\shortauthors}{Bando, Mita, Hiwada, Sano, Suzuki, Nakanishi, Tomida, Kajihara, Kaneko, Taki, Miyamoto, Yoshida, Shiozawa}

\begin{abstract}
  When key-value (KV) stores use SSDs for storing a large number of items, oftentimes they also require large in-memory data structures including indices and caches to be traversed to reduce IOs.
  This paper considers offloading most of such data structures from the costly host DRAM to secondary memory whose latency is in the microsecond range, an order of magnitude longer than those of DIMM-mounted persistent memory and currently available CXL memory devices.
  While emerging microsecond-latency memory, such as one based on flash memory, is likely to cost much less than DRAM, it can significantly slow down pointer-chasing on those in-memory data structures of SSD-based KV stores if naively employed, although its impact has not been well studied.
  This paper analyzes and evaluates the impact of microsecond-level memory latency on the throughput of SSD-based KV operations.
  Our analysis finds that a well-known latency-hiding technique of software prefetching for long-latency memory from user-level threads is effective for SSD-based KV stores.
  The novelty of our analysis lies in modeling how the interplay between prefetching and IO affects performance, from which we derive an equation that well explains the throughput degradation due to long memory latency.
  The model tells us that the presence of IO in KV operations significantly enhances their tolerance to memory latency, and the throughput degradation is expected to be small even if the memory latency extends to a few microseconds, leading to a finding that SSD-based KV stores can be made latency-tolerant without devising new techniques for microsecond-latency memory.
  To confirm this through experiments, we design a microbenchmark as well as modify existing SSD-based KV stores so that they issue prefetches for long-latency memory from user-level threads, and run them while placing most of in-memory data structures on FPGA-based memory with adjustable microsecond latency.
  The results demonstrate that their KV operation throughputs for varying memory latency can be well explained by our model, and the modified KV stores achieve near-DRAM throughputs for up to a memory latency of around 5 microseconds.
  This suggests the possibility that SSD-based KV stores involving latency-sensitive in-memory data traversal can use microsecond-latency memory as a cost-effective alternative to the host DRAM.
\end{abstract}



\received{April 2025}
\received[revised]{July 2025}
\received[accepted]{August 2025}

\maketitle

\newcommand\myeq[1]{Equation~#1}
\newcommand\myfig[1]{Figure~#1}

\newcommand\singlethreadsymbol{\textcolor[HTML]{7f7f7f}{$\boldsymbol{-\!\cdot\!\cdot-}$}}
\newcommand\multithreadsymbol{\textcolor[HTML]{e377c2}{$\boldsymbol{\cdot\!\cdot\!\cdot\!\cdot\!\cdot}$}}
\newcommand\prefetchlimitsymbol{\textcolor[HTML]{8c564b}{\rule[0.8mm]{7mm}{0.8pt}}}
\newcommand\fullmodelsymbol{\textcolor[HTML]{17becf}{$\boldsymbol{-\!\cdot\!-}$}}
\newcommand\maskonlysymbol{\textcolor[HTML]{9467bd}{$\boldsymbol{-\!-\!-}$}}

\newcommand\observation[2]{\vspace{1mm}\noindent\textbf{Observation O#1:} #2}

\newcommand\memorylatency{L_\mathrm{mem}}
\newcommand\dramlatency{L_\mathrm{DRAM}}
\newcommand\hidablelatency{\memorylatency^\ast}
\newcommand\memoryoptime{T_\mathrm{mem}}
\newcommand\contextswitchtime{T_\mathrm{sw}}
\newcommand\iooptime{T_\mathrm{IO}}
\newcommand\iooptimepre{\iooptime^\mathrm{pre}}
\newcommand\iooptimepost{\iooptime^\mathrm{post}}
\newcommand\iolatency{L_\mathrm{IO}}
\newcommand\waittime{T_\mathrm{wait}}
\newcommand\throughput{\Theta}
\newcommand\memop{\mathrm{mem}}
\newcommand\throughputmem{\Theta_\memop}
\newcommand\perfsingle{\mathrm{single}}
\newcommand\perfmulti{\mathrm{multi}}
\newcommand\perfworst{\mathrm{mask}}
\newcommand\perfbest{\mathrm{best}}
\newcommand\perfrandom{\mathrm{prob}}

\newcommand\usec{$\mu$sec}

\newcommand\perfrandomupdated{\mathrm{rev}}
\newcommand\perffull{\mathrm{extended}}
\newcommand\memorysize{A_\mathrm{mem}}
\newcommand\iosize{A_\mathrm{IO}}
\newcommand\memorybandwidth{B_\mathrm{mem}}
\newcommand\iobandwidth{B_\mathrm{IO}}
\newcommand\ioiops{R_\mathrm{IO}}
\newcommand\offloadratio{\rho}
\newcommand\evictratio{\varepsilon}

\section{Introduction}

\begin{figure}[t]
  \includegraphics[trim=0 220 740 0,clip,height=40mm]{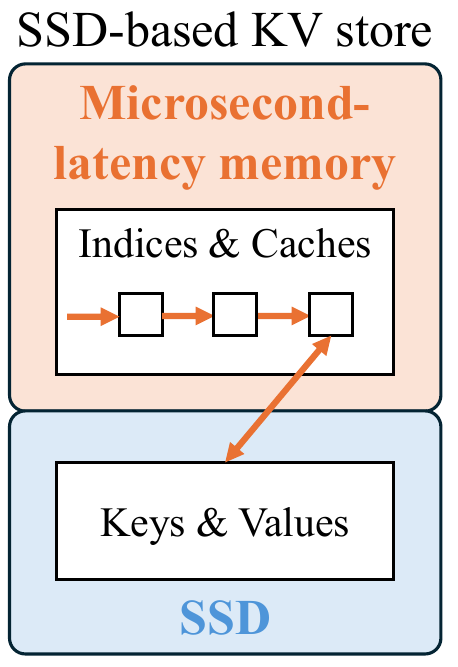}
  \hspace{1mm}
  \includegraphics[trim=20 270 540 10,clip,height=33mm]{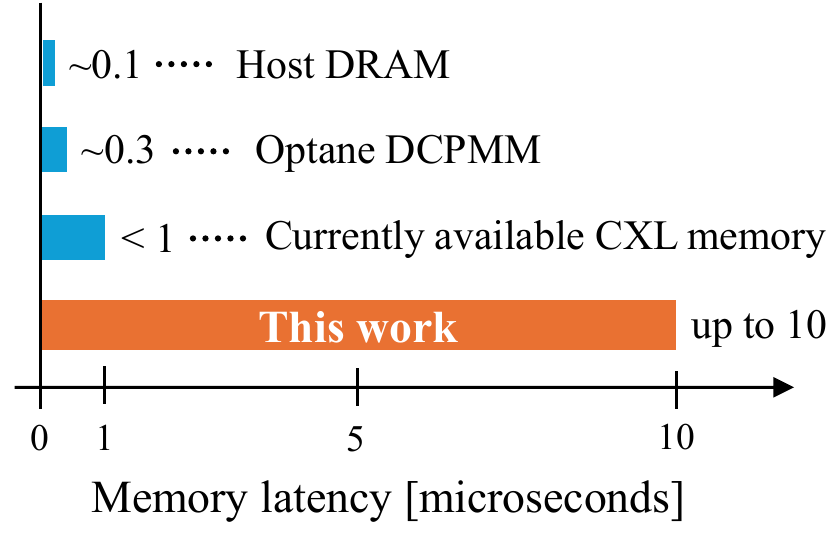}
  \mbox{(a) \hspace{33mm} (b) \hspace{15mm}}
  \caption{
    (a) SSD-based KV stores often perform pointer-chasing on large in-memory indices and caches in addition to SSD access.
    We offload these in-memory data structures from the host DRAM to much slower, microsecond-latency memory.
    (b) The memory latency we study is in the microsecond range, an order of magnitude longer than those of currently available memory devices.
    }
  \label{fig:teaser}
\end{figure}

Key-value (KV) stores play a vital role in various data center services \cite{FastKVStores,KVSFlashSurvey,KVWorkloadsFacebook,AmazonDynamoDB}.
As shown in \myfig{\ref{fig:teaser}}(a),
some KV stores use SSDs for storing a large number of items beyond the host DRAM capacity, and use in-memory data structures including indices and caches to minimize the number of time-consuming SSD accesses (i.e., IOs) while searching for items \cite{KVSFlashSurvey}.
The size of these in-memory data structures often grows as the database size increases, which can dominate the host DRAM usage \cite{IndeXY,FlashKVSSCM} (80--96\% of the memory footprint in our evaluation).
Since having large DRAM is costly, it would be beneficial to offload these large in-memory data structures to lower-cost secondary memory such as DIMM-mounted persistent memory (e.g., Intel Optane DCPMM) and memory based on Compute Express Link (CXL) \cite{CXL}.
While the bandwidth of secondary memory may be made to meet application requirements by using multiple memory devices, their longer latency can have a negative performance impact, as has been reported for memory devices with a latency of hundreds of nanoseconds \cite{IntelOptaneDCPMM,Pond}.
The issue is likely to be further exacerbated as emerging CXL devices as well as future memory devices may introduce even longer, microsecond-level latency if they employ lower-cost memory media including flash memory \cite{LongLatencyLowerCost,DissectCXLMemory}.
Naive adoption of such long-latency memory as a replacement for DRAM will significantly degrade the KV operation throughput (measured in operations per second, or ops/sec), given that those in-memory data structures typically require pointer-chasing (e.g., tree/linked-list traversal), which is sensitive to memory latency \cite{PointerChase1,PointerChase2,CoroBase,MosaicDB}.

This paper analyzes and evaluates the performance impact of offloading in-memory data structures of SSD-based KV stores from the host DRAM to much slower, microsecond-latency secondary memory.
Note that the memory latency we study in this paper is an order of magnitude longer than those of currently available devices as shown in \myfig{\ref{fig:teaser}}(b).
There is a spectrum of how secondary memory may be utilized, and one could consider more in-memory-oriented solutions of replacing SSDs by secondary memory \cite{Viper}.
But in this paper, we are interested in lower-cost solutions of replacing the host DRAM, which can account for half of the server cost \cite{Pond}.
In particular, this paper focuses on offloading of entire indices and caches, as this is a latency-sensitive base-case scenario where, in contrast to partial offloading, every pointer dereference for indices and caches is performed on secondary memory.
Through analysis and evaluation, we show that SSD-based KV stores can be made tolerant to microsecond-level memory latency using well-known techniques for hiding memory latency: software prefetching of data on microsecond-latency memory from user-level threads \cite{SoftwarePrefetching,KillerMicroseconds,UserLevelThreads}.
That is, multiple lightweight threads running in the user space process independent KV operations where each thread issues a prefetch to move the desired data to the CPU cache before it actually uses it.
To the best of our knowledge, it has not been reported in the literature that SSD-based KV stores with their in-memory indices and caches placed on microsecond-latency memory can still achieve KV throughputs that are close to those when the entire in-memory data is placed on the host DRAM.
Our analysis and evaluation show that this is possible without having to devise new techniques.

The novelty of our analysis lies in modeling how the presence of IOs makes SSD-based KV stores more latency-tolerant than they would be without IOs.
Prior work has experimentally shown the performance impact of microsecond-latency memory in in-memory settings (i.e., operations consist of memory accesses only), one of whose important observations is that
it is generally difficult for latency-sensitive applications to approach DRAM performance if the latency extends to a microsecond level \cite{KillerMicroseconds}.
We extend this prior work by considering operations consisting of memory accesses {\it as well as IOs}, and we study them experimentally {\it as well as theoretically}.
We introduce a generic operation model of SSD-based KV stores, and
derive an equation describing the expected KV throughput as a function of memory latency, leading to a key observation that the presence of IO significantly enhances latency-tolerance.
We show that simply adding the IO processing times to the throughput equation only partially explains this latency-tolerance enhancement.
Our theoretical analysis reveals how the interplay between memory prefetching and IO leads to further gain in latency-tolerance.

We validate our analysis through experiments.
As currently-available memory devices do not have microsecond-level latency, we use FPGA-based CXL memory devices designed to introduce user-specified latency.
We first run a microbenchmark that repeats memory accesses and IOs, and show that our throughput equation aligns well with microbenchmark performance.
We then show our analysis also applies to some existing SSD-based KV stores if they incorporate the above-mentioned latency-hiding techniques.
We take Aerospike \cite{Aerospike}, RocksDB \cite{RocksDB}, and CacheLib \cite{CacheLib} as representative SSD-based KV stores, and modify them so that they run user-level threads and issue software prefetches before they access microsecond-latency memory.
We demonstrate that our model still well explains how the throughputs of these modified KV stores behave as memory latency increases.

In summary, we make the following contributions.
\begin{itemize}
  \item Analysis: We model how the interplay between prefetching and IO affects the throughput of SSD-based KV stores,
  leading to a finding that SSD-based KV stores can be made latency-tolerant without devising new techniques for microsecond-latency memory.
  \item Implementation: We design a microbenchmark\footnote{Available at \url{https://github.com/ybandy/cxlkvs}}
  as well as modify some existing SSD-based KV stores (Aerospike\footnote{Available at \url{https://github.com/ybandy/aerospike-server}}, RocksDB\footnote{Available at \url{https://github.com/ybandy/rocksdb}}, and CacheLib\footnote{Available at \url{https://github.com/ybandy/CacheLib}}),
  so they run user-level threads and issue prefetches before they access microsecond-latency memory.
  \item Evaluation: By running the microbenchmark and the modified KV stores on FPGA-based memory with adjustable microsecond latency, we demonstrate that their operation throughputs can be well explained by our model, confirming that SSD-based KV stores can more readily be made latency-tolerant thanks to the presence of IO.
\end{itemize}
We believe these contributions provide novel insights into the performance of SSD-based KV stores on microsecond-latency memory that existing memory-only analysis does not provide.
Our evaluation shows that the modified KV stores achieve near-DRAM throughputs, suggesting the possibility that SSD-based KV stores involving latency-sensitive in-memory traversal can use microsecond-latency memory as a cost-effective alternative to the host DRAM.

\section{Objective and Observations}
Our goal is to show that it is possible for SSD-based KV stores to achieve near-DRAM performance even if most of their in-memory data structures are offloaded to microsecond-latency memory.
We make the following observations to reach our goal.
\begin{enumerate}[label=\textbf{O\arabic*:}]
\item Even with prefetching, data traversal slows down on microsecond-latency memory (Section~\ref{sec:memory_only_model}, reconfirmation of \cite{KillerMicroseconds})
\item IO significantly reduces this slowdown and makes prefetching more effective in theory (Section~\ref{sec:mem_and_io_model}) \label{obs:mem_and_io}
\item The throughput model based on O2 well explains microbenchmark performance, validating O2 in practice (Section~\ref{sec:microbench}).
\item The throughput model also agrees with some SSD-based KV stores in single-core, read-dominant cases, suggesting that IO makes it easier for SSD-based KV stores to become latency-tolerant (Section~\ref{sec:results_single_core}).
\item Latency-tolerance does not deteriorate by having other factors that slow down throughputs, including cache and lock contentions in multicore execution, write operations, and background workers (Section~\ref{sec:multicore}).
\end{enumerate}
Putting these observations together, our analysis and experiments show that SSD-based KV stores can be made tolerant to microsecond memory latency.

\section{Analysis}
\label{sec:analysis}

In this section, operation throughputs of SSD-based KV stores involving in-memory data traversal
are analyzed using a simplified model of a KV operation: a sequence of (long-latency) memory accesses followed by an IO.
Based on the model, we derive equations describing how throughputs depend on memory latency.

The key to (and the novelty of) our analysis is the presence of IOs.
With absence of IOs (i.e., in the memory-only case), performance impacts of microsecond-level memory latency have been studied in the past.
Cho et al. showed that (1) software prefetching could potentially hide microsecond latency and achieve throughputs that matched those of DRAM, but that (2) the full potential could not be attained due to a CPU hardware limitation on prefetching \cite{KillerMicroseconds}.
We will first review these two observations in the memory-only case and model these behaviors with equations in Section~\ref{sec:memory_only_model}.
We then move on to the memory-and-IO case in Section~\ref{sec:mem_and_io_model} and show that the presence of IOs ease the prefetch limitation, thereby boosting latency-tolerance.
In other words, our analysis shows that the issue observed by Cho et al. (their finding (2) above) can be mitigated by IOs without requiring to relax the CPU hardware limitation.

For tractable analysis, as well as for abstracting other factors away in order to concentrate on the impact of microsecond memory latency, we make the following assumptions.
We deal with execution on a single CPU core: multiple threads divide time and no two threads run at the same time.
We do not model overheads of threads: using more threads will not slow down each thread.
We do not model specifics or inner workings of secondary memory and SSD devices, nor do we consider bandwidth limits of secondary memory and SSDs: we simply look at the average memory access latency and IO processing times that a CPU core experiences.
We assume that the CPU has unlimited cache capacity: prefetching will always lead to a cache hit.
These simplifications keep the model generic and easy to interpret (some of them will be relaxed later in Section~\ref{sec:model_extension}).
The model still well explains actual performance of SSD-based KV stores.

In this section, we mainly consider the reciprocal of a throughput, that is, the time a CPU core spends (either by doing meaningful processing or by waiting) per operation, as this will simplify equations.
Table~\ref{tab:symbols} summarizes the symbols used in this paper, along with the ranges of their typical values as well as their example values used for illustration.
\begin{table}[t]
  \caption{Definition of Symbols}
  \label{tab:symbols}
  \centering
  \begin{tabular}{llll}
    \toprule
    Symbol                & Definition                 & Value range     & Example value \\
    \midrule
    $\throughput$         & KV operation throughput    & -               & -    \\
    $\memorylatency$      & Memory latency             & 1 -- 10 \usec    & -    \\
    $\memoryoptime$       & Memory suboperation time   & O(0.1) \usec    & 0.1 \usec \\
    $\iooptimepre$        & Pre-IO suboperation time   & O(1) \usec      & 4   \usec \\
    $\iooptimepost$       & Post-IO suboperation time  & O(1) \usec      & 3   \usec \\
    $\contextswitchtime$  & Context switch time        & O(0.1) \usec    & 0.05 \usec\\ 
    $N$                   & \# of threads              & -               & -    \\
    $P$                   & Prefetch queue depth       & O(10)           & 10   \\
    $M$                   & \# of memory accesses      & O(10)           & 10   \\
  \bottomrule
  \end{tabular}
\end{table}

\subsection{Memory-Only Model}
\label{sec:memory_only_model}

\begin{figure}[t]
  \centering
  \includegraphics[width=0.7\linewidth, trim=120 420 120 20, clip]{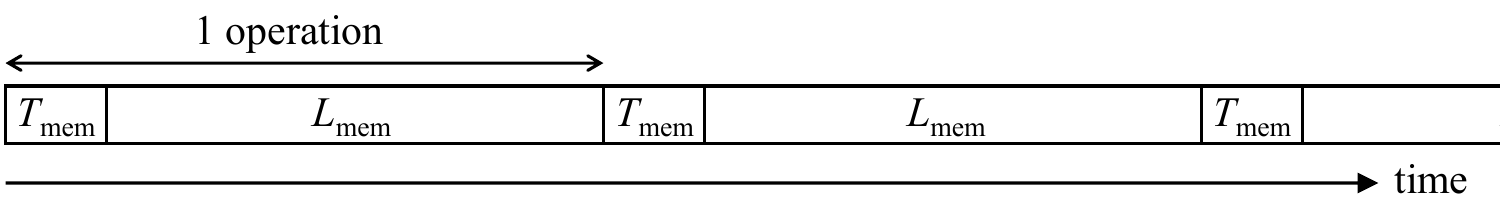}
  \caption{Memory-only operation model}
  \label{fig:model_memory_only_baseline}
\end{figure}

Here we derive throughput equations for memory-only model without IOs, which theoretically explain the empirical observations made in \cite{KillerMicroseconds}.
We consider the situation where a CPU core keeps executing memory accesses and some associated computations as shown in \myfig{\ref{fig:model_memory_only_baseline}}.
We let $\memorylatency$ denote the memory latency, and have $\memoryoptime$ absorb all the time a CPU core spends before it requests next data from memory.
In the memory-only model, one operation consists of one pair of computation and memory access for simplicity, but a sequence of these pairs will model in-memory data traversal in the memory-and-IO model in Section~\ref{sec:mem_and_io_model}.

\begin{figure}[t]
  \centering
  \includegraphics[width=0.7\linewidth, trim=0 330 560 0, clip]{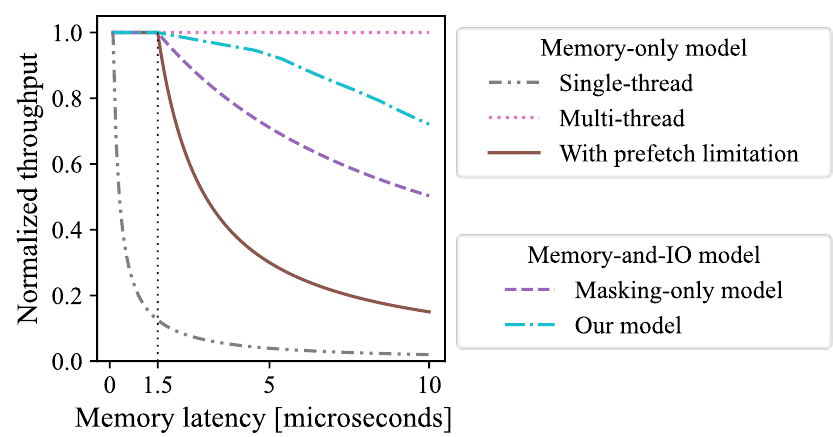}
  \caption{Normalized throughput calculated from various models using example values in Table~\ref{tab:symbols}}
  \label{fig:model}
\end{figure}

\subsubsection{Single-Threaded Case}

If we naively process these operations with a single thread, each operation takes $\memoryoptime + \memorylatency$ seconds, and the reciprocal throughput is given as
\begin{equation}
  \label{eqn:throughput_memop_baseline}
  \throughput_{\perfsingle}^{-1} = \memoryoptime + \memorylatency.
\end{equation}
Clearly, a longer latency $\memorylatency$ will lead to a lower throughput, as shown in \myfig{\ref{fig:model}} (\singlethreadsymbol).

\subsubsection{Multi-threaded Case}
\label{sec:mem_only_multi_threaded}

\begin{figure}[t]
  \includegraphics[width=0.9\linewidth, trim=10 10 0 50, clip]{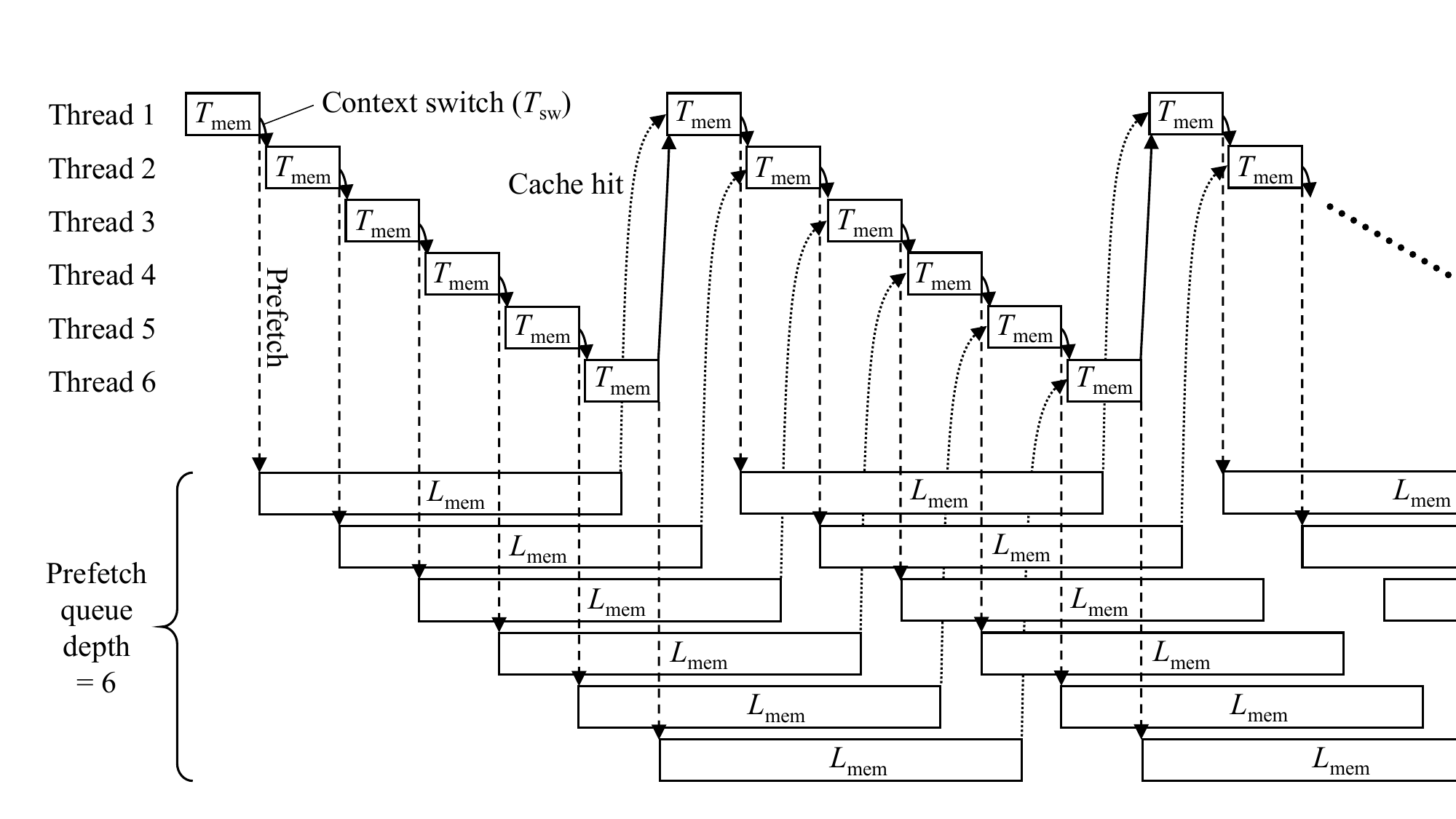}
  \caption{Memory-only operation model using multiple threads with a large prefetch queue depth}
  \label{fig:model_memory_only_multithread}
\end{figure}

Memory latency can be hidden by using multiple threads.
\myfig{\ref{fig:model_memory_only_multithread}} shows $N = 6$ threads responsible for independent operations involving repeated memory accesses.
When a thread (say Thread 1) needs to fetch data from long-latency memory, it issues a prefetch (dashed arrow) and yields to another thread (context switch to Thread 2, solid arrow).
While the data is being prefetched to the CPU cache, another thread can resume its own operation, which also performs a prefetch and context switch.
If the number of threads is large enough,
the data will be on the CPU cache by the time the control returns to the thread that issued a prefetch for the data (dotted arrow).
When the thread performs a memory load, it does not see the memory latency.

The reciprocal throughput in this case is given as
\begin{equation}
  \label{eqn:throughput_memop}
  \throughput_{\perfmulti}^{-1} = \max \left\{ \memoryoptime + \contextswitchtime, \; \frac{\memoryoptime + \memorylatency}{N} \right\}.
\end{equation}
The first term comes from the situation that has just been described above, where the time each thread spends on one operation becomes $\memoryoptime + \contextswitchtime$.
The second term is due to the Little's Law \cite{LittlesLaw}, stating that the reciprocal throughput is the length of one operation, which is $\memoryoptime + \memorylatency$ as shown in \myfig{\ref{fig:model_memory_only_baseline}}, divided by the number $N$ of threads.
No matter how many threads we have to decrease the second term, the CPU core has to spend $\memoryoptime + \contextswitchtime$ per operation as in the first term, and thus the maximum of the two.

\begin{figure}[t]
  \includegraphics[width=0.9\linewidth, trim=10 150 0 50, clip]{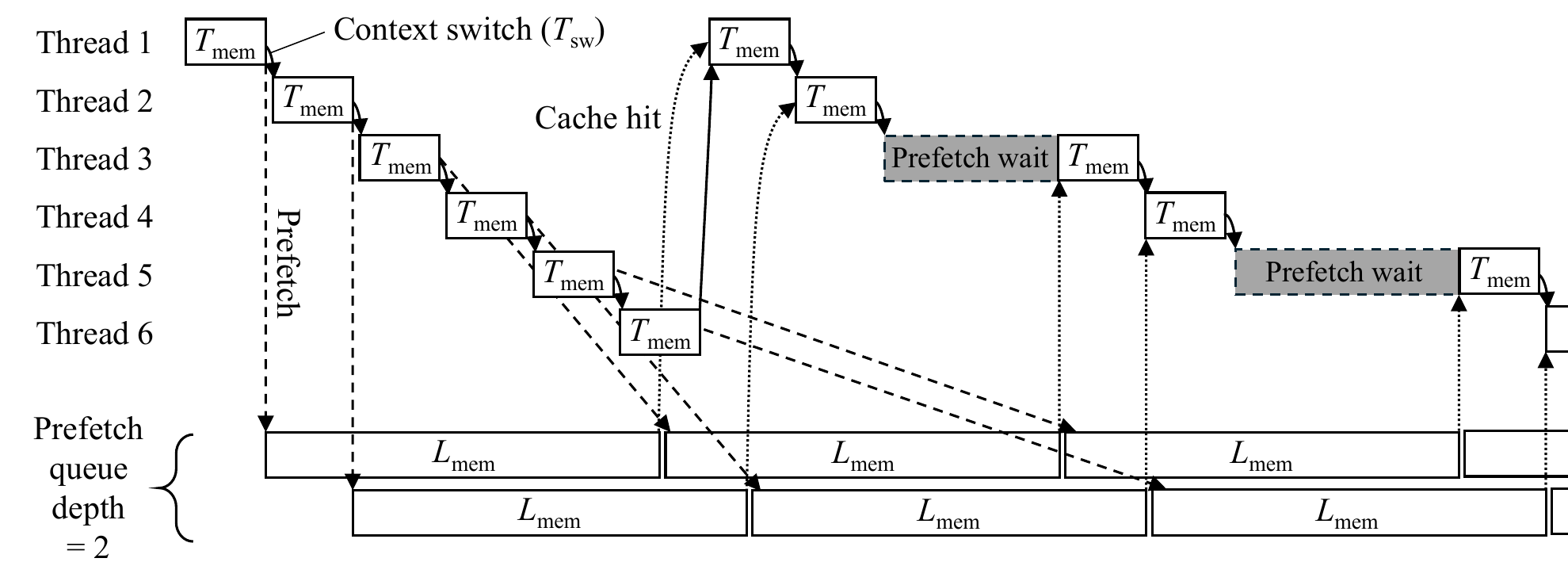}
  \caption{Memory-only operation model using multiple threads with a small prefetch queue depth}
  \label{fig:model_memory_only_prefetch_limit}
\end{figure}

\myeq{\ref{eqn:throughput_memop}} indicates that any arbitrarily long latency can be tolerated, at least in theory, by using a large number $N$ of threads to diminish the second term.
Then the throughput becomes $\throughput_{\perfmulti} = 1/(\memoryoptime + \contextswitchtime)$, which is constant as shown in \myfig{\ref{fig:model}} (\multithreadsymbol).

In order for this throughput to be better than the single-threaded case in \myeq{\ref{eqn:throughput_memop_baseline}}, we need to have a short context switch time $\contextswitchtime \ll \memorylatency$.
Conventional kernel-level threads are not effective as they have $\contextswitchtime \approx 1$ \usec{} or more.
Thus, \textit{user-level threads} \cite{UserLevelThreads}, able to switch contexts much more quickly ($\contextswitchtime \approx 0.1$ \usec) by working in the user space, are used.
Since user-level threads are our default choice, we simply call them ``threads'' unless otherwise noted.

\subsubsection{Multi-Threaded Case with Prefetch Limitation}

Cho et al. \cite{KillerMicroseconds} also identified an issue that could keep the approach described above from being effective: the depth $P$ of the prefetch queue per CPU core was limited to around $P = 10$, meaning that having more than $P$ threads did not improve performance.
For ease of illustration, let us assume we have only $P = 2$ in \myfig{\ref{fig:model_memory_only_prefetch_limit}}.
Since prefetches cannot start immediately (oblique dashed arrows), threads have to wait for the prefetches they issued to complete, wasting the CPU time (gray bars).
The throughput is determined by the Little's Law applied to the memory latency $\memorylatency$ and parallelism $P$.
As this imposes an additional limit to \myeq{\ref{eqn:throughput_memop}}, we have a full expression of the reciprocal throughput as
\begin{equation}
  \label{eqn:throughput_memop_full}
  \throughputmem^{-1} = \max \left\{ \memoryoptime + \contextswitchtime, \; \frac{\memoryoptime + \memorylatency}{N}, \; \frac{\memorylatency}{P} \right\}.
\end{equation}
Depending on the CPU hardware implementation, prefetch wait times may occur at different timings than depicted in \myfig{\ref{fig:model_memory_only_prefetch_limit}}, or prefetches can even be dropped \cite{FetchMe}.
In any case, when the prefetch queue is full, the subsequent load will incur a cache miss.
This limits the throughput, and \myeq{\ref{eqn:throughput_memop_full}} still holds.
No matter how many threads $N$ are used to minimize the second term of \myeq{\ref{eqn:throughput_memop_full}}, the third term deteriorates the throughput, resulting in the curve shown in \myfig{\ref{fig:model}} (\prefetchlimitsymbol).
By equating the first and the third terms, the latency $\hidablelatency$ beyond which the throughput deteriorates is given as
\begin{equation}
  \label{eqn:memory_latency_limit_full}
  \hidablelatency = P(\memoryoptime + \contextswitchtime).
\end{equation}
With the example values in Table~\ref{tab:symbols}, 
$\hidablelatency = 10 \times (0.1 + 0.05) = 1.5$ \usec, which shows why microsecond-latency memory deteriorates throughputs.
The issue does not usually manifest itself when the memory latency is in the sub-microsecond range.

\observation{1}{
Even with prefetching, data traversal slows down on microsecond-latency memory.
}

\begin{figure}[t]
  \centering
  \includegraphics[width=0.7\linewidth, trim=30 420 220 20, clip]{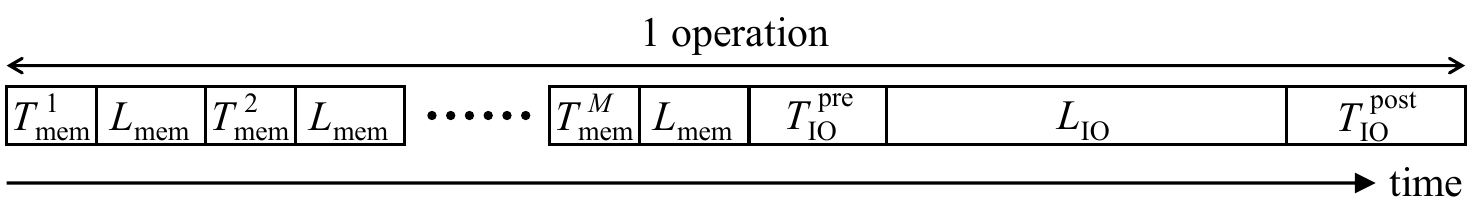}
  \caption{Memory-and-IO operation model}
  \label{fig:model_full_naive}
\end{figure}

\subsection{Memory-and-IO Model}
\label{sec:mem_and_io_model}

Now we extend our model to include IOs, and show that IO processing times hide memory latency and boost latency-tolerance.

One operation of an SSD-based KV store is modeled as a sequence of memory accesses followed by one IO as shown in \myfig{\ref{fig:model_full_naive}} (this configuration will be relaxed in Section~\ref{sec:model_extension}).
The number of memory accesses per operation is denoted by $M$ and there are $M$ associated computations each spending $\memoryoptime$ seconds (the superscripts in the figure are there to make it clear there are $M$ computations).
Since IOs also have long latency, we process them asynchronously.
Namely, one IO has three parts, (1) a pre-IO suboperation spending $\iooptimepre$ seconds including the time to compute a storage address and to submit an IO request in a non-blocking way, (2) IO latency $\iolatency$, and (3) post-IO suboperation spending $\iooptimepost$ seconds including the time to check IO completion, to copy the data to a buffer, and to use the data.
Note that the figure is not to scale: IO suboperations and latency may be much longer than those related to memory.

As we are interested in hiding both memory and IO latency, we skip the single-threaded case and discuss multi-threaded scenarios.

\begin{figure}[t]
    \centering
    \includegraphics[width=0.99\linewidth, trim=25 260 40 20, clip]{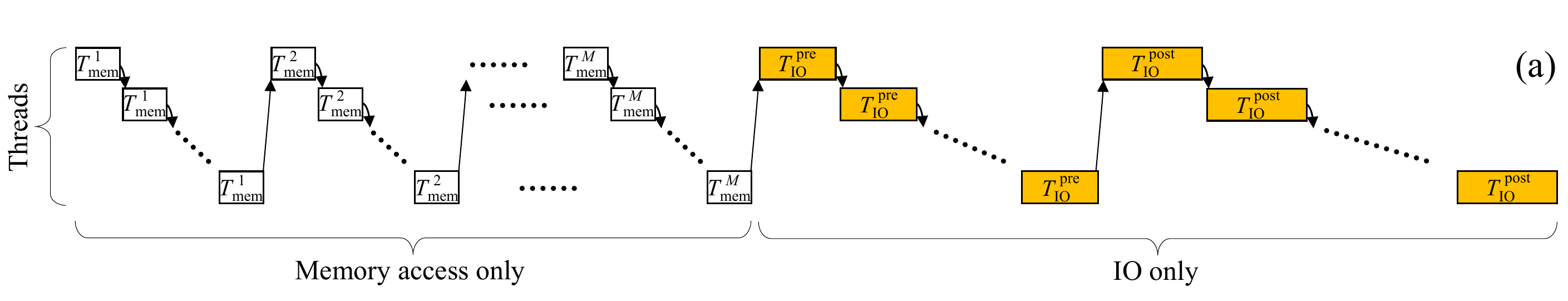}
    \includegraphics[width=0.99\linewidth, trim=25 340 40 20, clip]{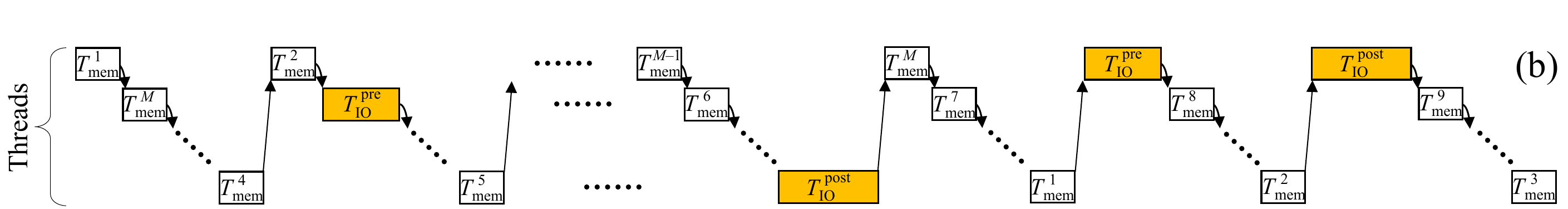}
    \caption{Memory-and-IO operation model with threads switching context both upon memory access and IO, where suboperations are (a) aligned and (b) misaligned across threads.
    IO suboperations are highlighted to clarify the difference between (a) and (b).}
    \label{fig:model_full}
\end{figure}

\subsubsection{Masking-Only Model}

A simple explanation of the performance impact of IO would be to add an extra processing time $E$ coming from IO to $M$ instances of the memory-only model as
\begin{equation}
  \label{eqn:throughput_mem_and_io_worst}
  \throughput_\perfworst^{-1} = M \throughputmem^{-1} + E,
\end{equation}
where $E$ is the total time a CPU core spends in processing IO as
\begin{equation}
  \label{eqn:total_io_times}
  E = \iooptimepre + \iooptimepost + 2\contextswitchtime,
\end{equation}
which is typically a few microseconds.
Here we assume that we have a sufficiently large number of threads to fully hide the IO latency $\iolatency$ as SSDs have deep queues.

According to \myeq{\ref{eqn:throughput_mem_and_io_worst}}, the offset $E$ makes the degradation in the overall throughput $\throughput_\perfworst$ smaller than that in the memory-only throughput $\throughputmem$ alone, as shown in \myfig{\ref{fig:model}} (\maskonlysymbol{} over \prefetchlimitsymbol).
For example, consider an extreme case where $E$ is very large.
Then, small change in $\throughputmem$ will not affect $\throughput_\perfworst$ much.
In other words, the IO-related processing times $E$ mask the performance impact of long-latency memory $M \throughputmem^{-1}$, and we call this the \textit{masking-only model}.

The masking effect increases latency-tolerance, but not to the point where we approach DRAM performance.
In our example of \myfig{\ref{fig:model}} (\maskonlysymbol), the masking-only model predicts 29\% throughput degradation at a memory latency of 5 \usec, which is considerable.
The reason of this degradation is because the few-microsecond offset $E$ coming from IO processing times is not large enough to make the memory access time $M \throughputmem^{-1}$ negligible.
When the memory latency is in the microsecond range, we have $M \throughputmem^{-1} = M\memorylatency/P$ from \myeq{\ref{eqn:throughput_memop_full}}.
Continuing with the same example of $P = M = 10$ as above, this reduces to $M \throughputmem^{-1} = \memorylatency$.
Hence, $M \throughputmem^{-1}$ and $E$ in \myeq{\ref{eqn:throughput_mem_and_io_worst}} are both a few microseconds and are comparable to each other.
While it may be tempting to think that the impact of memory latency is small in the face of IO processing times, that is not the case in general if the memory latency is in the microsecond range.

\subsubsection{Our Model}
\label{sec:our_model}

Now we identify the issue in the masking-only model, and show that IO can further enhance latency-tolerance by deriving a better model.
The key is whether the available prefetch depth is used efficiently.
To understand this, \myfig{\ref{fig:model_full}}(a) shows the case where suboperations are aligned, meaning that all the threads perform $\memoryoptime^1$ successively, then move on to $\memoryoptime^2$, and so on.
Since the left-hand side of this figure consists of memory accesses alone, and the right-hand side IOs alone, the reciprocal throughput is the sum of those for the both sides, as in \myeq{\ref{eqn:throughput_mem_and_io_worst}}.
Therefore, now we know that the masking-only model represents the scenario where IO does not serve to hide memory latency.

The observation above indicates that if we mix memory and IO suboperations by ``misaligning'' threads as shown in \myfig{\ref{fig:model_full}}(b), we have a better utilization of the available prefetch depth.
In the best-case scenario, the throughput cap due to prefetch depth only applies to the entire sequence, resulting in a reciprocal throughput as follows.
\begin{equation}
  \label{eqn:throughput_mem_and_io_best}
  \throughput_\perfbest^{-1}
  = \max \left\{ M (\memoryoptime + \contextswitchtime) + E, \;
  \frac{M \memorylatency}{P} \right\}.
\end{equation}
Then, by comparing the two terms in \myeq{\ref{eqn:throughput_mem_and_io_best}}, the maximum latency $\hidablelatency$ that does not deteriorate the throughput is
\begin{equation}
  \label{eqn:memory_latency_limit_with_io_best}
  \hidablelatency = P(\memoryoptime + \contextswitchtime) + \frac{PE}{M}.
\end{equation}
This is longer than that of the memory-only case of \myeq{\ref{eqn:memory_latency_limit_full}} by $PE/M$.
With the example values in Table~\ref{tab:symbols}, we have $PE/M = 7.1$ \usec{} and thus $\hidablelatency = 8.6$ \usec, which is much longer than without IO (1.5 \usec).

\begin{figure}[t]
  \centering
  \includegraphics[height=50mm, trim= 30 100 500 30, clip]{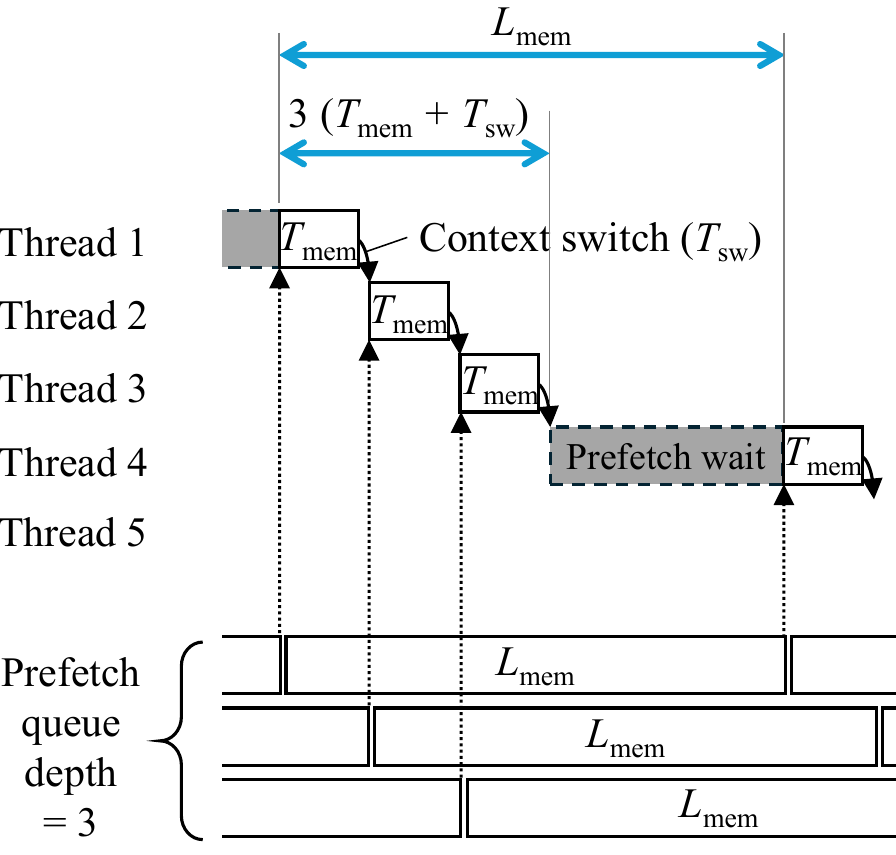}
  \hspace{1mm}
  \includegraphics[height=50mm, trim=130 100 500 30, clip]{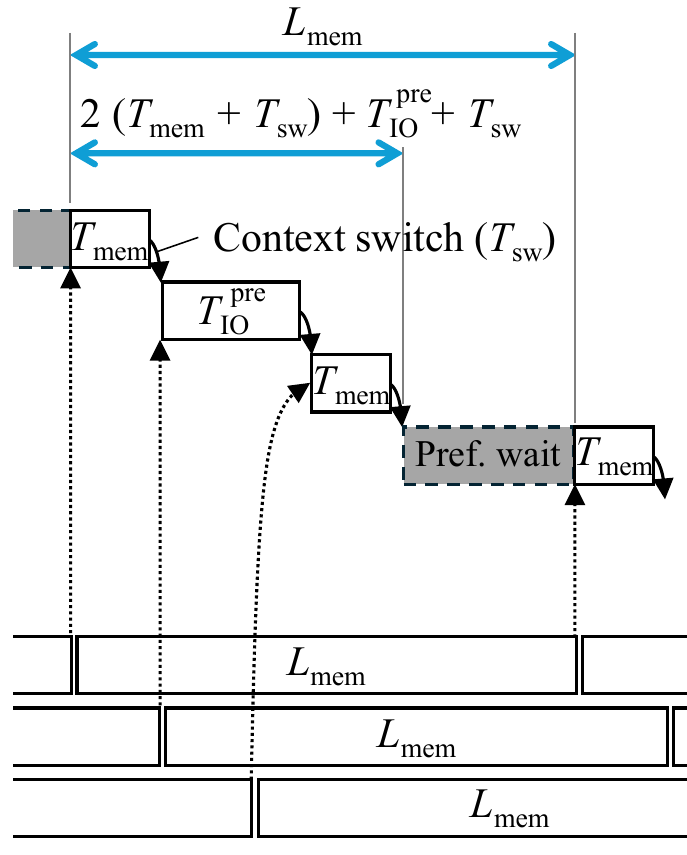}
  \hspace{1mm}
  \includegraphics[height=50mm, trim=130 100 500 30, clip]{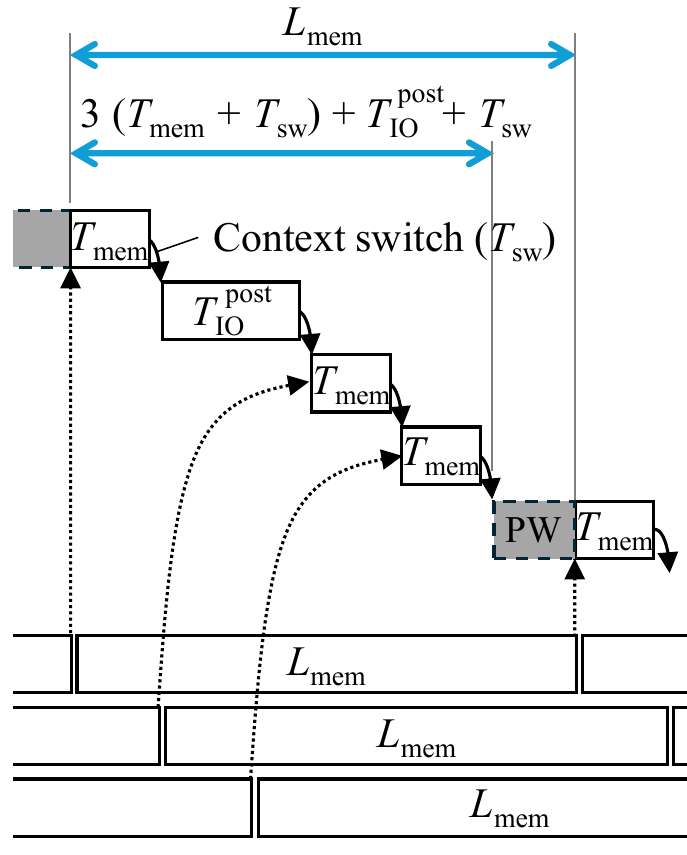}
  \mbox{\hspace{9mm} (a) \hspace{37.3mm} (b) \hspace{37.3mm} (c)}
  \caption{Prefetch wait time (a) with no IO suboperation, (b) with a pre-IO, and (c) with a post-IO suboperation}
  \label{fig:prefetch_wait}
\end{figure}

In reality, it is not practical to control execution timing of threads relative to each other.
Fortunately, timing will not be aligned as in \myfig{\ref{fig:model_full}}(a) either, but will be mostly random.
So far, we have considered the number $M$ of memory accesses a constant, but in practice, it varies from operation to operation because traversal ends whenever the searched item is found.
While we will still consider the {\it average} number of memory accesses to be $M$, its variance naturally misaligns thread timing, albeit not optimally in terms of latency-tolerance.
We are interested in the throughput in this practical scenario.
Our analysis strategy is to model prefetch wait times probabilistically and compute the expected wait time per operation.

We first see how one pre- and post-IO suboperations reduce prefetch wait times in \myfig{\ref{fig:prefetch_wait}}.
As shown in \myfig{\ref{fig:prefetch_wait}}(a), in prefetch depth-limited scenarios, a prefetch wait time happens every $P$ memory accesses ($P = 3$ in the figure), and the wait time will be $\memorylatency - P (\memoryoptime + \contextswitchtime)$.
However, as shown in \myfig{\ref{fig:prefetch_wait}}(b), if one pre-IO suboperation replaces one of $P$ memory accesses, the wait time will be reduced by $\iooptimepre - \memoryoptime$, as a $\memoryoptime$-second suboperation is replaced by a $\iooptimepre$-second one.
\myfig{\ref{fig:prefetch_wait}}(c) shows the case of a post-IO suboperation.
As a post-IO suboperation waits for an IO but not for a prefetch, it will defer the wait time by one thread (Thread 5 experiences the wait time rather than Thread 4), and the wait time will be reduced by $\iooptimepost + \contextswitchtime$.

We generalize these observations to cases with more IO suboperations.
We consider a sequence of $P + k$ suboperations consisting of $j$ pre-IO, $k$ post-IO, and $P - j$ memory suboperations.
In other words, we have $P$ memory suboperations in the beginning, but $j$ out of $P$ are replaced by pre-IOs, and additionally $k$ post-IOs are inserted, making a sequence of length $P + k$.
Extending the wait time reduction arguments described above to multiple IO suboperations, the wait time the $(P + k)$-th thread will experience is given as
\begin{equation}
  \label{eqn:wait_time}
  \waittime(j, k) = \max \{0, \, \memorylatency - P (\memoryoptime + \contextswitchtime)
  - j (\iooptimepre - \memoryoptime) - k (\iooptimepost + \contextswitchtime)\},
\end{equation}
where $\max\{0, \cdot\}$ is because the wait time is non-negative.

Next, we compute the probability that this suboperation sequence happens.
To make the analysis tractable, we make the following simplification:
we assume memory and IO suboperations to occur independently and identically.
Namely, a suboperation that a CPU core processes next will be one of the following.
\begin{itemize}
  \item Memory suboperation $\memoryoptime$ with a probability of $M/(M+2)$,
  \item Pre-IO suboperation $\iooptimepre$ with a probability of $1/(M+2)$, or
  \item Post-IO suboperation $\iooptimepost$ with a probability of $1/(M+2)$.
\end{itemize}
Then, the probability $p(j, k)$ that the above-mentioned sequence with $j$ pre-IO and $k$ post-IO suboperations happens is
\begin{equation}
  \label{eqn:sequence_probability}
    p(j, k) = \frac{(P+k)!}{(P-j)! \, j! \, k!} \left(\frac{M}{M+2}\right)^{P-j} \left(\frac{1}{M+2}\right)^{j+k}.
\end{equation}

Finally, we compute the expected prefetch wait time.
Suppose we have many ($i = 1, 2, \cdots, n$) of these sequences each of which is of length $P + k_i$ and contains $(j_i, k_i)$ IO suboperations.
In the long run, the average wait time per suboperation will be
\begin{equation}
  \frac{\waittime(j_1, k_1) + \waittime(j_2, k_2) + \cdots + \waittime(j_n, k_n)}{(P + k_1) + (P + k_2) + \cdots + (P + k_n)}.
\end{equation}
When $n$ is large, the numerator and denominator behave as independent Gaussian random variables (Central Limit Theorem), and hence the expected per-suboperation wait time $\waittime^\mathrm{subop}$ can be approximated by the ratio of the expectations (denoted by $\mathbb{E}[\cdot]$) of the numerator and denominator \cite{RatioDistributions} as 
\begin{equation}
  \label{eqn:expected_wait_time}
  \waittime^\mathrm{subop} \approx \frac{\mathbb{E}_{j, k}[\waittime(j, k)]}{\mathbb{E}_{j, k}[P + k]}
  = \frac{\sum_{j = 0}^P \sum_{k = 0}^\infty p(j, k) \, \waittime(j, k)}{\sum_{j = 0}^P \sum_{k = 0}^\infty p(j, k) \, (P + k)}.
\end{equation}
To compute these expectations, one does not have to deal with infinite summation ($k \rightarrow \infty$) in practice, since $p(j, k)$ quickly vanishes for large $k$.
As one operation has $M+2$ suboperations, the expected reciprocal throughput according to our probabilistic formulation is
\begin{equation}
  \label{eqn:throughput_mem_and_io_random}
  \throughput_\perfrandom^{-1}
  = M (\memoryoptime + \contextswitchtime) + E + (M+2) \, \waittime^\mathrm{subop}.
\end{equation}
The throughput degradation predicted by this model is plotted in \myfig{\ref{fig:model}} (\fullmodelsymbol).
The degradation is much smaller, 7\% at a memory latency of 5 \usec, compared with 29\% of the masking-only model (\maskonlysymbol).
This is because IOs interleave with prefetches and alleviate the slowdown coming from the prefetch limitation.

\observation{2}{
IO significantly reduces the slowdown due to long memory latency and makes prefetching more effective.
}

\subsubsection{Model Extension}
\label{sec:model_extension}
Our memory-and-IO model, which has assumed one IO to follow $M$ memory accesses, can be extended to cases where multiple IOs appear anywhere in the sequence.
If one operation has $S$ IOs on average, we can split it into $S$ smaller operations each with $M/S$ memory accesses and one IO, and use \myeq{\ref{eqn:throughput_mem_and_io_random}} to calculate the expected throughput.
As this is $S$-fold overcounting, we divide it by $S$.
Since scale does not matter in discussing throughput degradation, we consider $M$ to be a per-IO value in what follows.
Our model applies to any order of suboperations as our formulation only assumes probabilities of suboperations.

\begin{table}[t]
  \caption{System Parameters}
  \label{tab:system_params}
  \centering
  \begin{tabular}{lll}
    \toprule
    Symbol                & Definition                        & Example value \\
    \midrule
    $\memorysize$         & Memory access (cacheline) size    & 64 bytes     \\
    $\memorybandwidth$    & Maximum memory bandwidth          & 10 GB/sec    \\
    $\iosize$             & SSD access (IO) size              & O(1) kB      \\
    $\iobandwidth$        & Maximum SSD bandwidth             & 10 GB/sec    \\
    $\ioiops$             & Maximum SSD random access         & 2.2 MIOPS    \\
    $\offloadratio$       & Offloading ratio of indices and caches   & 1            \\
    $\evictratio$         & Premature CPU cache eviction ratio          & $\approx 0$  \\
  \bottomrule
  \end{tabular}
\end{table}

For completeness, we further extend our model by relaxing some of the simplifications described in the beginning of Section~\ref{sec:analysis}.
Table~\ref{tab:system_params} lists additional symbols that will come into play.
The right column of the table shows example values for these system parameters, which we use in our evaluation in Section~\ref{sec:eval}.
The reciprocal throughput according to our extended model is given as follows.
\begin{equation}
  \label{eqn:full_model}
  \throughput_{\perffull}^{-1} = \max \left\{
    \throughput_{\perfrandomupdated}^{-1}, \;
    \frac{\iosize}{\iobandwidth}, \;
    \frac{1}{\ioiops}
    \right\},
\end{equation}
where $\iosize$ is the average IO access size (in bytes), and $\iobandwidth$ and $\ioiops$ are the maximum bandwidth (bytes/sec) and random access performance (IOPS) of the SSD, respectively.
$\throughput_{\perfrandomupdated}$ is a revised version of the probabilistic throughput model of \myeq{\ref{eqn:throughput_mem_and_io_random}} taking into account the memory bandwidth and CPU cache capacity limits as well as memory tiering of DRAM and secondary memory.
$\throughput_{\perfrandomupdated}$ can be obtained through a similar derivation to that in Section~\ref{sec:our_model} with the following two modifications.
First, we replace the memory latency $\memorylatency$ in Equation~\ref{eqn:wait_time} as
\begin{equation}
  \label{eqn:memory_latency_extension}
  \memorylatency \leftarrow \max \left\{
    \offloadratio \, \memorylatency + (1-\offloadratio) \, \dramlatency, \;
    (P-j) \, \frac{\memorysize}{\memorybandwidth}
    \right\},
\end{equation}
where $\offloadratio$ is the offloading ratio to the secondary memory, $\dramlatency$ is the DRAM latency, $\memorysize$ is the memory access size (in bytes), and $\memorybandwidth$ is the maximum memory bandwidth (bytes/sec).
The offloading ratio $\offloadratio$ is in terms of access frequency, such that the average memory latency in the long run becomes a linear interpolation of the two types of memory latencies as in the first term of \myeq{\ref{eqn:memory_latency_extension}}.
For example, $\offloadratio = 0.7$ means that 70\% of accesses to indices and caches go to the secondary memory and the rest to the DRAM.
The second term of \myeq{\ref{eqn:memory_latency_extension}} is because a suboperation sequence containing
$P-j$ memory suboperations takes at least $(P-j)\memorysize/\memorybandwidth$ seconds due to the memory bandwidth limit.

The second modification in the derivation relates to the CPU cache capacity limit.
If the capacity is small, some of prefetched data may be evicted from the cache before it is referenced.
We call this probability a premature CPU cache eviction ratio $\evictratio$.
In the original derivation, we assume one memory suboperation to occur with a probability of $M/(M+2)$.
We split this into two cases depending on whether it is executed before or after the prefetched data is evicted.
\begin{itemize}
  \item Pre-eviction memory suboperation with a probability of $(1-\evictratio) M/(M+2)$, and
  \item Post-eviction memory suboperation with a probability of $\evictratio M/(M+2)$,
\end{itemize}
where the former can be treated in the same way as the original memory suboperation, while the latter behaves in the same way as a post-IO suboperation except that it takes $\memorylatency$ seconds instead of $\iooptimepost$.
Thus, we can follow the same derivation as in Equations~\ref{eqn:sequence_probability}--\ref{eqn:expected_wait_time} by using different probabilities and suboperation times.

The simplification we do not address here is the overhead of threads.
In practice, using more threads leads to a slowdown in throughputs, mainly caused by the increased cost of context switching and CPU cache contention.
However, both of them are difficult to theoretically model, as they depend on many other factors such as the thread implementation, CPU cache configuration, memory footprint, memory address assignment, and memory access patterns.
Fortunately, the impact of the thread overhead seems limited as we will see later.

\section{Evaluation}
\label{sec:eval}

\begin{table}[t]
  \caption{Computational Environment}
  \label{tab:environment}
  \centering
  \begin{tabular}{ll}
    \toprule
    Part       & Specifications \\
    \midrule
    CPU        & 2 of Intel Xeon Gold 6430 (32 cores/CPU, 2.10 GHz) \\
    DRAM       & DDR5 4800 MHz 512 GB (32 GB $\times$ 8 ch./CPU) \\
    CXL        & 2 of Intel Agilex 7 FPGA I-Series Dev. Kit (128 GB in total) \\
    CXL        & 1 of LR-LINK CXL Memory Expander (64 GB) \\
    SSD        & 4 of Intel Optane 900P 480 GB NVMe (1.92 TB in total) \\
    OS         & Ubuntu 22.04.5 LTS, Linux kernel 6.8.0 \\
  \bottomrule
  \end{tabular}
\end{table}

We conduct performance evaluation to support our analysis.
We first run a microbenchmark to validate our throughput model (Section~\ref{sec:microbench}).
We then demonstrate that SSD-based KV stores can be made tolerant to microsecond-level memory latency thanks to the presence of IOs (Section~\ref{sec:kv_stores}).

Our computational environment is summarized in Table~\ref{tab:environment}.
As currently-available memory devices do not have microsecond-level latency,
we use FPGA-based CXL memory.
Two FPGA boards are each equipped with 64-GB DRAM, and we implement an FPGA circuit so that the memory latency can be adjusted to an arbitrary length \cite{XLGPU}.
Additionally, we use a commercially-available CXL memory expander equipped with DRAM.
While the measured latency of this device is around 300 nanoseconds, it allows us to check the validity of our model at that specific, sub-microsecond latency.
All of these CXL devices are attached via PCIe links, and appear as CPU-less NUMA nodes.
Hyper-Threading and hardware prefetching are disabled so that performance is more predictable and interpretable, as is also done in other works~\cite{KillerMicroseconds,DemystifyCXL,MosaicDB}.

The system parameters of our computational environment are shown in the right column of Table~\ref{tab:system_params}.
They are chosen so as not to obscure the latency-dependence of KV stores.
We use multiple devices so that the combined bandwidths $\memorybandwidth$ and $\iobandwidth$ and random access performance $\ioiops$ do not limit the KV throughputs.
We offload the entire KV indices and caches to secondary memory ($\offloadratio = 1$) so that the KV throughputs always depend on the predetermined microsecond-level latency.
The CPU cache size is large enough (60 MB L3) so as not to limit the KV throughputs.
As we will see later, the premature CPU cache eviction ratio $\evictratio$ is close to zero.
Therefore, our simplified model of \myeq{\ref{eqn:throughput_mem_and_io_random}} suffices to describe our experimental conditions.
Nonetheless, we validate our extended model of \myeq{\ref{eqn:full_model}} by changing the system parameters to deliberately violate our simplifying assumptions in Section~\ref{sec:extended_model_validation}.
To this end, our FPGA-based memory is designed to be able to throttle the memory bandwidth.

\begin{figure}[t]
  \centering
  \includegraphics[width=0.7\linewidth, trim=30 280 340 50, clip]{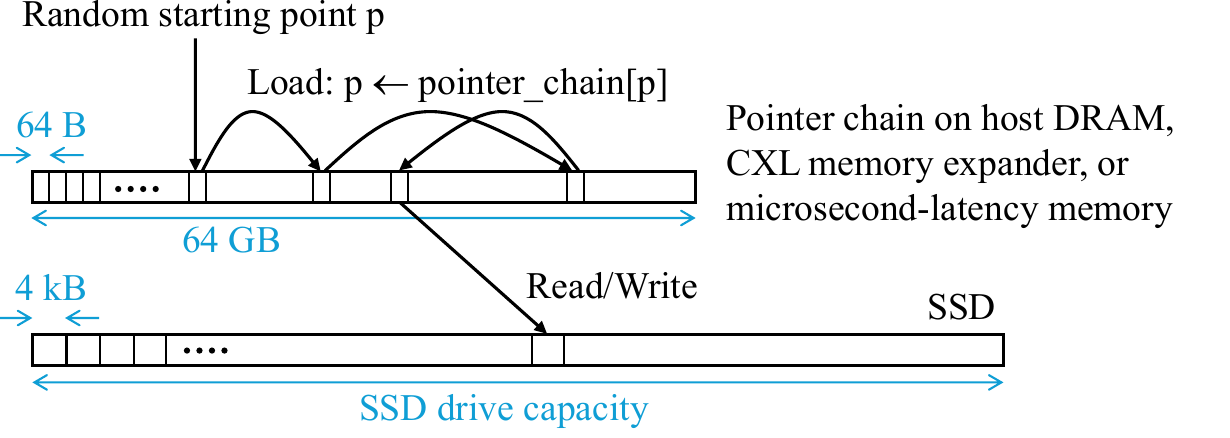}
  \caption{One operation of the microbenchmark}
  \label{fig:microbench_diagram}
\end{figure}

\subsection{Microbenchmark}
\label{sec:microbench}

We design a microbenchmark that performs exactly the operation modeled in Section~\ref{sec:analysis}, and confirm that observed throughputs align well with the throughput equation of \myeq{\ref{eqn:throughput_mem_and_io_random}}.

\subsubsection{Implementation}

In our microbenchmark, each operation consists of $M$ successive memory accesses followed by pre-IO and post-IO suboperations as shown in \myfig{\ref{fig:model_full_naive}}, and multiple threads keep processing independent operations as in \myfig{\ref{fig:model_full}}.

\myfig{\ref{fig:microbench_diagram}} illustrates one operation in more detail.
Memory suboperations perform pointer chasing, so that a next access depends on its previous one, simulating latency-sensitive data traversal.
We use Argobots \cite{Argobots} for user-level threads, and each time a thread performs a memory suboperation, it issues a prefetch for the next pointer address and yields to another thread.
We place the entire pointer chain on any one of the host DRAM, CXL memory expander, or microsecond-latency memory.
The chain elements are permuted to minimize spatial locality.
The size of the pointer chain is 64 GB (1 billion of 64-byte cacheline-sized pointers).
Each operation picks a random starting point and performs pointer chasing $M$ times.
After pointer chasing, an SSD is accessed based on the final pointer (i.e., random read or write).
We access an SSD as a block device to avoid performance irregularity due to a file system.
Asynchronous IO is implemented with io\_uring \cite{IOuring}.

\subsubsection{Experimental Conditions}

We run the microbenchmark on a single core for 1,404 ($= 4 \times 3 \times 3 \times 3 \times 13$) combinations of different parameter values and memory latencies as follows.
\begin{itemize}
  \item The number of memory accesses $M = \{1, 5, 10, 15\}$
  \item Memory suboperation time $\memoryoptime = \{0.10, 0.12, 0.14\}$ \usec
  \item Pre-IO suboperation time $\iooptimepre = \{1.5, 2.5, 3.5\}$ \usec
  \item Post-IO suboperation time $\iooptimepost = \{0.2, 1.2, 2.2\}$ \usec
  \item Memory latency $\memorylatency = \{0.1, 0.3, 0.5, 1, 2, 3, \cdots, 10\}$ \usec
\end{itemize}

The variations in $\memoryoptime$ are created with a spin loop by calling \texttt{pause} 1, 3, and 5 times.
From the host DRAM execution, we estimate one call of \texttt{pause} consumes 10 nsec with an offset of 90 nsec, thus $\memoryoptime = 90 + 10 \times 5 = 140$ nsec if we call \texttt{pause} 5 times, for instance.

The variations in $\iooptimepre$ and $\iooptimepost$ are created by adding extra 1 or 2 \usec{} to the IO submission and completion checking times.
The IO submission and completion checking times are estimated to be 1.5 and 0.2 \usec, respectively, by running an IO-only benchmark (i.e., $M = 0$).

For each parameter combination, we evaluate the throughput dependence on memory latency.
When the pointer chain is placed on the host DRAM, the memory latency is about 0.1 \usec.
When on the CXL memory expander, it is about 0.3 \usec.
When on the microsecond-latency memory, we change its latency from 0.5 to 10 \usec{} (0.5 \usec{} is the minimum latency of our FPGA-based memory).
For each latency, we optimize the number of threads: namely, we try different numbers of threads (on a single core) and report the highest throughput.
Since we observe similar throughputs for read IOs and write IOs, we report read IO results.

\begin{figure}[t]
  \centering
  \includegraphics[trim=0 420 500 0, clip, width=0.7\linewidth]{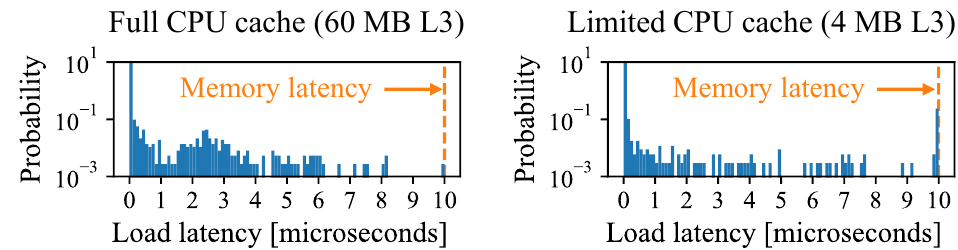}
  \mbox{\hspace{10mm} (a) \hspace{45mm} (b)}
  \caption{Probability distribution of load latency}
  \label{fig:load_latency}
\end{figure}

We confirm it is rare that prefetched data gets evicted from the CPU cache before it is referenced during pointer chasing.
Using PEBS \cite{PEBS} via \texttt{perf mem} command, we measure latencies experienced by load instructions for accessing the secondary memory.
As shown in \myfig{\ref{fig:load_latency}}(a), most (note the log scale) of the load latencies are close to zero, indicating cache hits.
Some loads wait for a few microseconds due to late prefetches caused by the prefetch queue limit.
Only a small fraction ($\evictratio < 0.0005$) of loads suffer premature cache eviction and wait for as long as the set memory latency (10 \usec{} in the figure).
By drastically reducing the L3 cache size from 60 MB to 4 MB using \texttt{resctrl} \cite{Resctrl}, we observe a non-negligible eviction ratio of $\evictratio \approx 0.05$ as shown in \myfig{\ref{fig:load_latency}}(b).

\begin{figure}[t]
  \centering
  \includegraphics[trim=0 90 410 20, clip, width=0.85\linewidth]{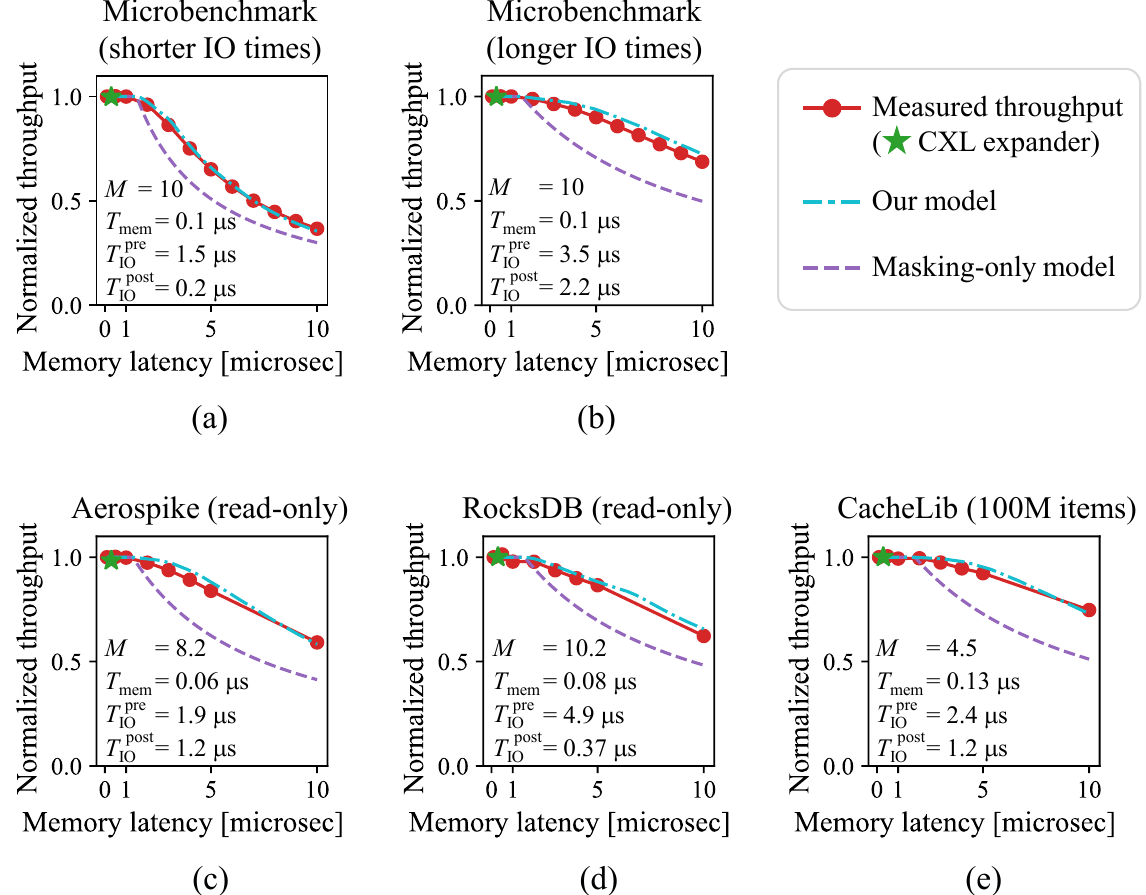}
  \caption{Normalized throughputs of (a)(b) the microbenchmark and of (c)(d)(e) the modified SSD-based KV stores on a single core, along with those predicted by the models}
\label{fig:single_core}
\end{figure}

\subsubsection{Results}
\myfig{\ref{fig:single_core}}(a)(b) shows throughputs for some representative parameter combinations.
The throughputs are normalized by that on DRAM to show how much throughput degradation occurs as the memory latency increases.
Along with the actual measurements, throughput degradations predicted by the masking-only model (\myeq{\ref{eqn:throughput_mem_and_io_worst}}) and by our probabilistic model (\myeq{\ref{eqn:throughput_mem_and_io_random}}) are shown.
The model throughputs are calculated from the same parameter values $(M, \memoryoptime, \iooptimepre, \iooptimepost)$ used to run the microbenchmark.
The other model parameters are estimated as $\contextswitchtime = 50$ nsec and $P = 12$ via \myeq{\ref{eqn:throughput_memop_full}} by running the microbenchmark with no IOs.

While the masking-only model underestimates performance for long memory latency, our model better approximates it.
In all the 1,404 parameter and latency combinations, the masking-only model underestimates the actual performance by up to 32.7\%, while our model is within $[-5.0\%, +6.8\%]$ of the actual measurements. This indicates that the latency-tolerance gained from IO cannot be solely explained by masking.
IO eases the prefetch limitation and allows prefetching to better hide latency.
By comparing the two plots in \myfig{\ref{fig:single_core}}(a)(b), we can also see that the latency-tolerance is better in (b) than in (a) because (b) has longer IO suboperations.
The improvement over the masking-only explanation is also larger in (b).
Additionally, these microbenchmark results indicate that we can achieve DRAM-equivalent throughputs using the commercial CXL memory expander as predicted by our model.

\observation{3}{
  The throughput model based on Observation O2 (IO enhances latency-tolerance) well explains microbenchmark performance, validating O2.
}

\begin{figure}[t]
  \centering
  \includegraphics[trim=0 90 400 20, clip, width=0.86\linewidth]{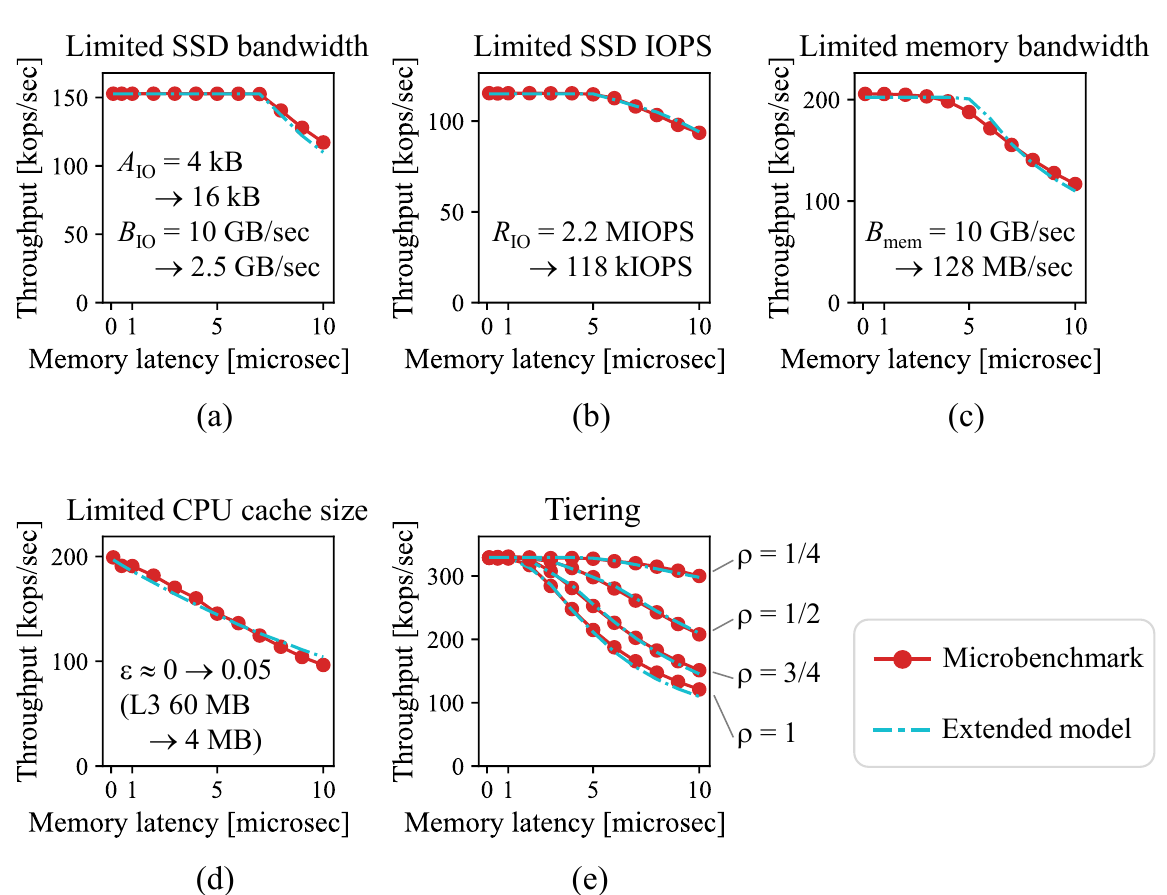}
  \caption{Throughputs of the microbenchmark and those predicted by the extended model under the conditions where other factors than memory latency affect performance}
\label{fig:extended_model}
\end{figure}

\subsubsection{Validation of Extended Model}
\label{sec:extended_model_validation}
The above results are obtained under the condition that the simplifications made in Section~\ref{sec:analysis} hold.
However, violation of them can have practical performance impacts, and we illustrate them in \myfig{\ref{fig:extended_model}}.
\myfig{\ref{fig:extended_model}}(a) shows an SSD bandwidth-limited scenario by increasing the access size $\iosize$ and by using only one SSD to decrease the bandwidth $\iobandwidth$.
Until the memory latency becomes too long to saturate the SSD bandwidth, the throughput is capped by the bandwidth and stays the same.
Similar behaviors are observed when the throughput is limited by the SSD random access performance $\ioiops$ (realized by using a slow SATA SSD) and by the memory bandwidth $\memorybandwidth$ (throttled by the FPGA) as shown in (b) and (c), respectively.
\myfig{\ref{fig:extended_model}}(d) shows a CPU cache size-limited scenario.
The latency-tolerance deteriorates since prefetching becomes ineffective.
\myfig{\ref{fig:extended_model}}(e) shows tiering where the pointer chain is split into $(1-\offloadratio):\offloadratio$ and they are placed on DRAM and the secondary memory, respectively.
As expected, we observe better latency-tolerance for smaller offload ratio $\offloadratio$.

In all of these scenarios, the observed throughputs are well explained by our extended model.

\begin{figure}[t]
  \centering
  \includegraphics[trim=0 120 210 0, clip, width=0.7\linewidth]{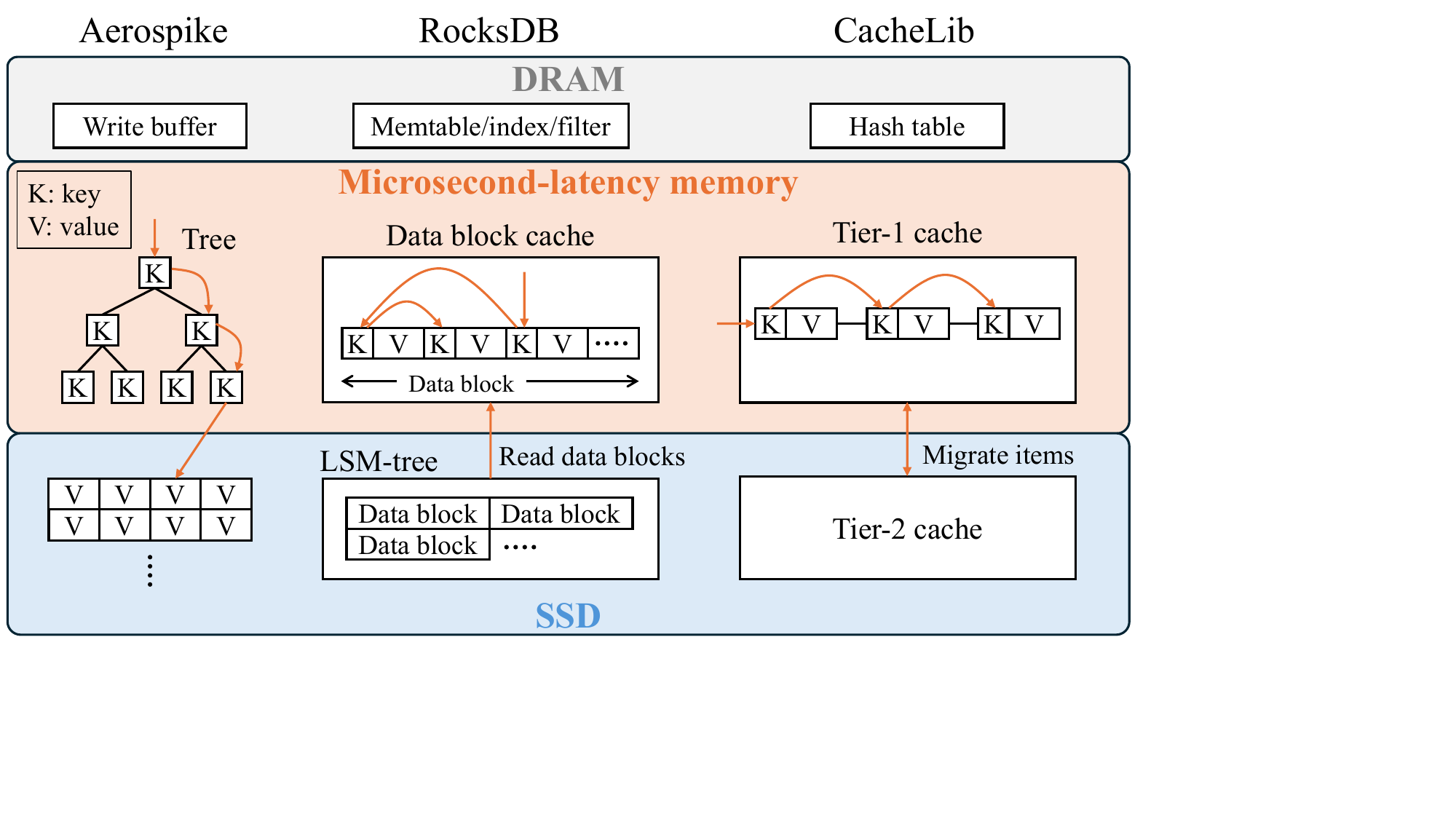}
  \caption{
    SSD-based KV stores performing pointer-chasing on large in-memory data structures in addition to accessing SSDs.
    We place these large data structures on microsecond-latency memory, while leaving other small data structures on the host DRAM.
    }
  \label{fig:data_structures}
\end{figure}

\subsection{KV Store Examples}
\label{sec:kv_stores}

Now we demonstrate that existing SSD-based KV stores can be made tolerant to memory latency with some modifications, and our analysis holds for these modified KV stores, indicating that their latency-tolerance is also enhanced by the presence of IOs.

To test with different SSD-based KV store designs, we take Aerospike \cite{Aerospike}, RocksDB \cite{RocksDB}, and CacheLib \cite{CacheLib} as examples.
As shown in \myfig{\ref{fig:data_structures}}, Aerospike traverses in-memory trees before accessing values on SSDs.
RocksDB fetches a data block from a log-structured merge-tree (LSM-tree) on SSDs and traverses sorted keys in the data block in an in-memory block cache.
CacheLib is a two-tier KV cache where the first, in-memory tier has linked items and LRU lists to be traversed while using SSDs as the second tier.
Even though these KV stores are different, they all involve in-memory data traversal and IOs modeled by our analysis.
As these in-memory indices and caches are the largest DRAM consumers in the respective SSD-based KV stores, we offload them in their entirety to secondary memory.
Offloading other data structures would have much less contribution to cost reduction, and they remain in the host DRAM as depicted at the top of \myfig{\ref{fig:data_structures}}.

Our extended model in Section~\ref{sec:model_extension} allows one KV operation to issue a varying number of IOs, covering RocksDB and CacheLib.
Beyond that, we do not need the model extension of \myeq{\ref{eqn:full_model}} since our system parameters in Table~\ref{tab:system_params} are chosen so that factors other than memory latency do not limit the KV throughputs.

\begin{table}[t]
  \caption{Code Modifications}
  \label{tab:code_modifications}
  \centering
  \small{
  \begin{tabular}{lll}
    \toprule
    Modification & Original code & Modified code \\
    \midrule
    \makecell[l]{Memory\\allocation}
      & index\_t* indices = malloc(size);
      & \makecell[l]{index\_t* indices = mmap($\cdots$, size, $\cdots$);\\
                     mbind(indices, size, $\cdots$, \&nodemask, $\cdots$);} \\
    \midrule
    \makecell[l]{Use of\\user-level\\threads}
      & \makecell[l]{Kernel-level thread function family\\
                     \textbullet \quad pthread\_create($\cdots$),\\
                     \textbullet \quad pthread\_join($\cdots$), etc.}
      & \makecell[l]{User-level thread equivalents\\
                     \textbullet \quad user\_level\_thread\_create($\cdots$),\\
                     \textbullet \quad user\_level\_thread\_join($\cdots$), etc.}\\
    \midrule
    \makecell[l]{Insertion\\of prefetch\\and yield}
      & index\_t val = indices[i];
      & \makecell[l]{\_\_builtin\_prefetch(indices + i);\\
                     user\_level\_thread\_yield();\\
                     index\_t val = indices[i];} \\
    \midrule
    \makecell[l]{Asynchro-\\nous IO}
      & pread(fd, buf, count, offset);
      & \makecell[l]{struct io\_uring\_sqe *sqe = io\_uring\_get\_sqe(\&ring);\\
                     io\_uring\_prep\_read(sqe, fd, buf, count, offset);\\
                     io\_uring\_submit(\&ring);\\
                     user\_level\_thread\_yield();\\
                     io\_uring\_peek\_cqe(\&ring, \&cqe);} \\
  \bottomrule
  \end{tabular}}
\end{table}

\subsubsection{Implementation}

We modify these KV stores so that they
\begin{itemize}
  \item allocate in-memory indices and caches on microsecond-latency memory,
  \item replace kernel-level threads with user-level threads,
  \item issue a prefetch and yield upon accessing indices and caches, and
  \item use an asynchronous IO interface.
\end{itemize}
Table~\ref{tab:code_modifications} exemplifies these modifications in C code.
They are written generically, and we refer the reader to the open-sourced modified code for complete modifications specific to each KV store.

As CXL memory devices (expander cards as well as our FPGA-based memory) appear as CPU-less NUMA nodes, memory can be allocated on them by specifying nodes in \texttt{nodemask} for \texttt{mbind}.

Kernel-level thread functions are replaced by user-level thread counterparts.
As with the microbenchmark, we use Argobots \cite{Argobots} as user-level threads for Aerospike.
For RocksDB, we adopt an existing modification \cite{Alibaba} that uses PhotonLibOS \cite{PhotonLibOS}.
Since CacheLib incorporates Folly library~\cite{Folly}, we use Folly Fibers.

To insert a prefetch and yield, we need to search the code for all accesses to secondary memory.
To this end, we repurpose Valgrind Memcheck memory error detector \cite{Valgrind}. 
By marking the indices and caches on the secondary memory as inaccessible and by running a KV workload, accesses to them are reported as illegal.
We insert a prefetch and yield before each detected memory access.

For asynchronous IO, we use io\_uring \cite{IOuring}.
A standard IO such as \texttt{pread} and \texttt{pwrite} is replaced by an IO submission (the first three lines of code in Table~\ref{tab:code_modifications}) and IO completion checking (last line).
We also insert a yield after an IO submission to hide IO latency.

The modified KV stores run faster than the original implementations in our evaluation.
When storing all of the in-memory data structures on the host DRAM, they show 1.2x higher throughputs than their original counterparts mainly thanks to the lightweight threads and IO.

\subsubsection{Experimental Conditions}
We use the benchmark tools accompanying the respective KV stores: Aerospike Benchmark for Aerospike, db\_bench for RocksDB, and CacheBench for CacheLib.
Table~\ref{tab:variation} summarizes experimental parameters for the modified KV stores.
Our default settings are shown in bold letters, many of which follow those of the benchmark tools.
We first run benchmarks using the default settings, and later make variations by changing each parameter as listed in the table.
Other parameters are chosen as follows.

\begin{table}[t]
  \caption{KV Store Settings}
  \label{tab:variation}
  \centering
  \begin{tabular}{lccc}
    \toprule
    Parameter      & Aerospike       & RocksDB        & CacheLib \\
    \midrule
    \# items       & \textbf{500M}   & \textbf{1B}    & \textbf{100M}, 400M \\
    \midrule
    Key size (bytes) & \textbf{20} & 10, \textbf{20}, 40 & [4, 8], \textbf{[8, 16]}, [16, 32] \\
    \midrule
    \multirow{3}{*}{Value size (bytes)}
                   & 1 k,            &  200,          & [100, 150], \\
                   & \textbf{1.5 k}, &  \textbf{400}, & \textbf{[200, 300]}, \\
                   & 2--2.5 k        &  800           & [300, 450] \\
    \midrule
    \multirow{2}{*}{Distribution}
                   & \textbf{Uniform},   &  \textbf{Zipf 0.99},  & \textbf{Gaussian}, \\
                   & Zipf 1.1            &  Zipf 0.8             & graph cache leader* \\
    \midrule
    Read:write     & \textbf{1:0}, 2:1, 1:1   &  \textbf{1:0}, 2:1, 1:1   & \textbf{2:1}, 1:1  \\
  \bottomrule
  \end{tabular}\\
  \small{* One of the workloads in CacheBench. We use its key distribution.}
\end{table}

\textbf{Aerospike:}
We set the default value size to 1.5 kB according to Aerospike Certification Tool \cite{AerospikeACT}.
By Aerospike's design, the size of each tree node is always 64 bytes regardless of the key size, thus we do not vary the key size.
By placing the trees (32 GB = 64 bytes $\times$ 500M) on secondary memory, the host DRAM usage is only 1.3 GB (96\% offload), which is mainly a write buffer.

\textbf{RocksDB:}
Since the block cache is effective for skewed key distributions,
we implement Zipfian distribution in db\_bench.
With a Zipf exponent of 0.99 and a block cache size of 32 GB, the block cache hit ratio is 67\%.
The other in-memory data structures consume 8 GB of the host DRAM, meaning that 80\% (32 out of 40 GB) of the memory consumption is offloaded to secondary memory.

\textbf{CacheLib:}
We set the value size to a few hundred bytes as observed in many web services \cite{WorkloadAnalysisKVS,Kangaroo}, and thus Small Object Cache \cite{CacheLib} is used as a tier-2 (SSD) cache.
We first run a relatively small workload encountering 100 million items with 8-GB tier-1 (host DRAM or secondary memory) and 32-GB tier-2 caches.
Since it can take a long time for the throughput of CacheLib to stabilize especially on a single core, 
running a small workload makes the warm up period short enough (under 6 hours) to allow us to iterate with varying numbers of CPU cores.
Then, we run a larger workload on 16 cores using 32-GB tier-1 and 128-GB tier-2 caches to deal with 400 million items.
When secondary memory is used, the remaining DRAM usage is 4.3 GB (65\% offload) and 7.8 GB (80\% offload) for the smaller and larger workloads, respectively, which are mostly attributed to the hash table.
Once the throughput is stabilized, the cache hit ratio is 82\% (34\% at tier-1 and 73\% at tier-2 upon tier-1 misses).
We further increase the tier-2 cache size to 512 GB later in Section~\ref{sec:cost}.

\subsubsection{Single-Core Results}
\label{sec:results_single_core}

We first demonstrate that the throughputs of the modified KV stores as functions of memory latency follow our analysis.
\myfig{\ref{fig:single_core}}(c)(d)(e) shows throughputs of the modified SSD-based KV stores for varying memory latency.
They are normalized by the throughput observed when the entire in-memory data structures are placed on the host DRAM.
As before, normalized throughputs according to the models are also shown.
The measured model parameters are shown in each plot.
They are measured by recording timestamps before and after prefetch yields and IO yields during host DRAM execution.
This is done separately from throughput measurements as the recording incurs some overheads.

As with the microbenchmark results, the actual performance is close to what our probabilistic model predicts, and is higher than the masking-only explanation.
Again, this indicates that IO eases the prefetch limitation and allows prefetching to better hide latency.

\observation{4}{
  The throughput model agrees with the modified SSD-based KV stores in single-core, read-dominant cases, suggesting that IO makes it easier for SSD-based KV stores to become latency-tolerant.
}

\begin{figure}[t]
  \centering
  \includegraphics[trim=0 330 510 0, clip, width=0.7\linewidth]{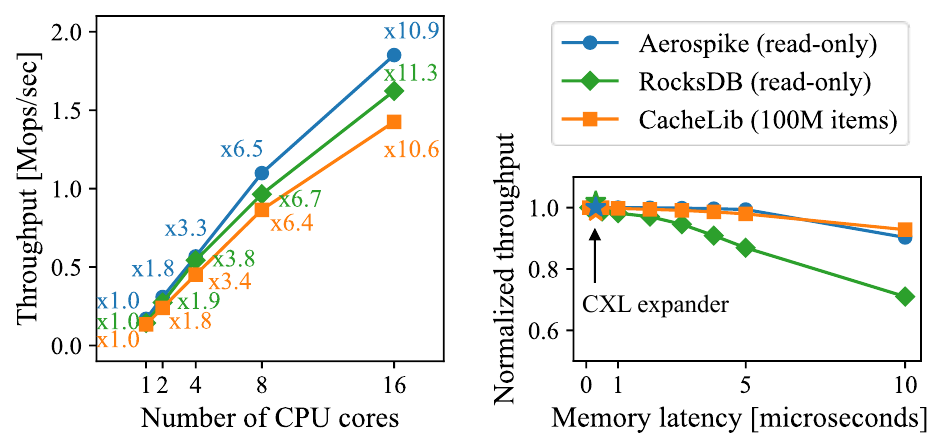}
  \mbox{\hspace{6mm} (a) \hspace{45mm} (b)}
  \caption{Multi-core throughputs of the modified SSD-based KV stores
  (a) for varying numbers of cores at a fixed memory latency of 5 usec, and
  (b) for varying memory latency at a fixed number (16) of cores.
  }
  \label{fig:multicore}
\end{figure}

\begin{figure}[t]
  \centering
  \includegraphics[trim=5 210 286 0, clip, width=\linewidth]{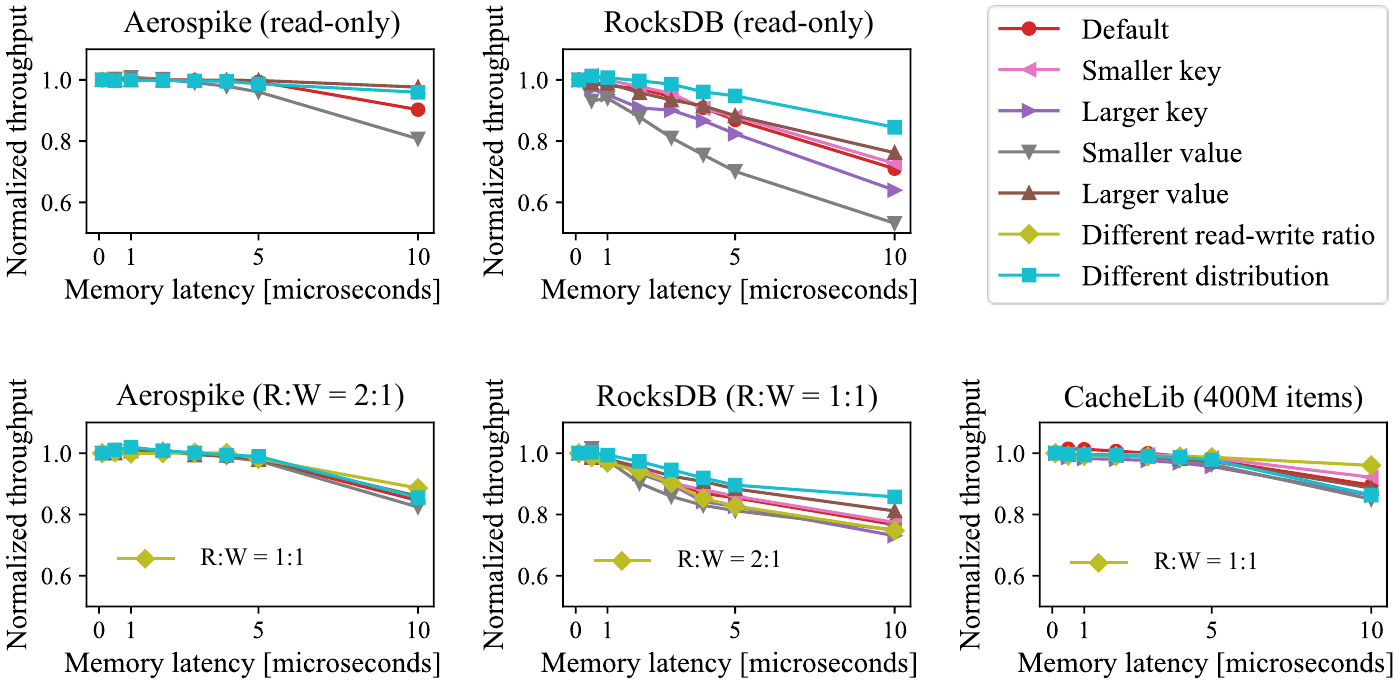}
  \caption{
    Normalized throughputs of the modified SSD-based KV stores with various settings
  }
  \label{fig:variation}
\end{figure}

\subsubsection{Multi-Core Results}
\label{sec:multicore}

We show that the latency-tolerance observed in the single-core execution does not deteriorate when the number of cores increases.
We run the same workload as in Section~\ref{sec:results_single_core} on 2, 4, 8, and 16 cores. 
For a given number of core and memory latency, we test different numbers of threads per core and pick the best throughput.
The throughputs of the modified KV stores with different numbers of cores are plotted in \myfig{\ref{fig:multicore}}(a).
Since plots of different memory latencies behave similarly, we only show those for a memory latency of 5 \usec.
The vertical axis shows raw throughputs, while the numbers in the plot show the throughputs normalized by that on a single core.
The throughput scales nicely as the number of cores increases, although it is sublinear (1.8--1.9x when the number of cores doubles).
The sublinearity likely comes from increased lock and CPU cache contentions.
This slowdown can favorably affect latency-tolerance as it masks throughput degradation due to memory latency.
As shown in \myfig{\ref{fig:multicore}}(b), with 16-core execution, RocksDB maintains the same level of latency-tolerance compared with the single-core case in \myfig{\ref{fig:single_core}}(d), while Aerospike and CacheLib see better tolerance with less than 2\% throughput degradation up to a memory latency of 5 \usec.

We also vary KV store settings as shown in Table~\ref{tab:variation} to see how they influence latency-tolerance.
The top and bottom rows of \myfig{\ref{fig:variation}} show read-only and write-mix cases, respectively.
For each case, we test smaller/larger key/value sizes, different key distributions, and for the write-mix case, different read-write ratios.
In most settings, we observe similar levels of latency-tolerance to that with the default settings.
Notable exceptions with worse tolerance are when we use smaller values in the read-only case.
This is because the slowdown coming from IO becomes smaller, making the impact of memory latency relatively larger.
The opposite effect occurs for RocksDB with less skewed distribution.
More block cache misses lead to more IOs, resulting in better latency-tolerance.
Write-mix settings make latency-tolerance less dependent on other factors as their impacts are masked by the increased burst SSD writes and the overhead of background workers for defragmentation and compaction.
The throughput degradation is 8\% at a memory latency of 5 \usec{} when averaged (geomean) over all of these measurements.

\begin{figure}[t]
  \centering
  \includegraphics[trim=0 390 560 10, clip, width=0.62\linewidth]{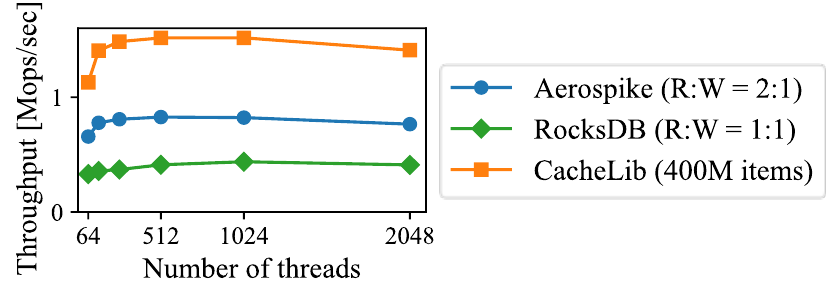}
  \caption{Throughput dependence of the modified SSD-based KV stores on the number of threads}
  \label{fig:stability}
\end{figure}

\myfig{\ref{fig:stability}} shows that throughput degradation due to the thread overhead is small.
Unless too many threads are used, the peak throughput is fairly stable across varying numbers of threads.

\begin{figure}[t]
  \centering
  \includegraphics[trim=0 390 300 0, clip, width=\linewidth]{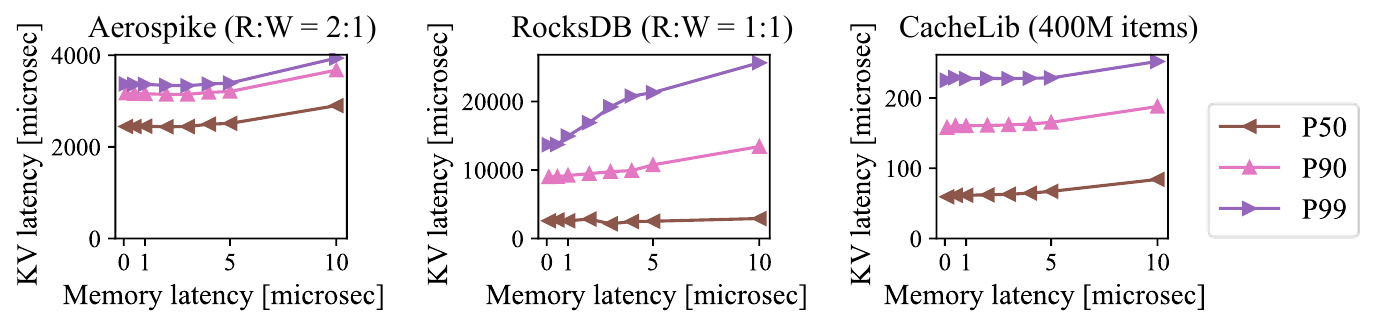}
  \caption{KV operation latency of the modified SSD-based KV stores}
\label{fig:latency}
\end{figure}

\myfig{\ref{fig:latency}} shows KV operation latency.
Longer memory latency leads to longer KV operation latency, but the impact is limited.

\observation{5}{
  Latency-tolerance does not deteriorate by having other factors that slow down throughputs, including cache and lock contentions in multicore execution, write operations, and background workers.
}

\section{Discussion}

By making Observations O1-5, our analysis and evaluation have found that SSD-based KV stores involving latency-sensitive in-memory traversal can be made tolerant to microsecond memory latency.
This suggests that one could use devices with low-cost memory media including CXL-based secondary memory using compressed DRAM \cite{CXLCompressedDRAM} and flash memory \cite{DissectCXLMemory}.
Such devices are still not readily available, and accurately modeling their expected performance is out of the scope of this paper.
However, we can make crude estimates of the system cost-performance improvement they would potentially gain, which we discuss below in Section~\ref{sec:cost}.
After that, we discuss future directions of our work in Section~\ref{sec:future}.

\subsection{Cost-Performance Estimates}
\label{sec:cost}

When part of the host DRAM is replaced by secondary memory, the cost-performance ratio (CPR) $r$ of the resultant system relative to the original DRAM-only system can be expressed as
\begin{equation}
  \label{eqn:cost}
  r = \frac{1 - d}{c b + (1 - c)},
\end{equation}
where $c$ (< 1) is the cost of DRAM being replaced by the secondary memory relative to the total server cost, $b$ (< 1) is the relative bit cost (dollars/GB) of the secondary memory to DRAM, and $d$ (< 1) is the throughput degradation caused by the use of the secondary memory.
A CPR $r$ exceeding 1 means cost-performance improvement.

Table~\ref{tab:cost} shows some example values for $b$ and $d$.
The bit cost $b$ of compressed DRAM comes from its typical compression ratio of 2-3x \cite{CXLCompressedDRAM}, which increases the effective capacity and thereby reduces the bit cost.
The bit cost of the flash-based memory device is based on low-latency SLC (single-level cell) flash.
The relative bit cost of SLC is 0.15 according to \cite{SLCcost2}.
We add some range to it by conservatively assuming that SLC is up to ten (rather than four) times more expensive than QLC (quad-level cell), where the relative bit cost of QLC is estimated to be 0.02 from retail prices of SSD and DRAM \cite{MemVsFlash}.

For throughput degradation $d$, since the latency of compressed DRAM is under 1 \usec{} \cite{CXLCompressedDRAM}, we refer to the 0-2\% degradation observed in our evaluations.
For flash-based memory, we assume a typical latency of 5 \usec.
We also simulate tail latency by configuring our FPGA to occasionally introduce longer latencies of 14 and 48 \usec{} at probabilities of 9.9\% and 0.1\%, respectively, which fit the relative latency distribution of a low-latency SSD \cite{LowLatencyFlash} under our FPGA design constraints.
With these settings, we conduct additional experiments and observe 2-19\% throughput degradation.

Given these numbers, Table~\ref{tab:cost} shows CPR values $r$ in a hypothetical scenario where DRAM accounts for half of the server cost \cite{Pond} and we replace 80\% of it with secondary memory (i.e., $c = 1/2 \times 0.8 = 0.4$).
These CPR values well exceed one, suggesting that these secondary memory devices will likely bring about cost-performance improvement.

\begin{table}[t]
  \caption{Example CPR parameters and values}
  \label{tab:cost}
  \centering
  \begin{tabular}{lrrr}
    \toprule
    Memory medium     & Bit cost $b$ & Degradation $d$ & CPR $r$ \\
    \midrule
    Compressed DRAM   & $1/3-1/2$   &  $0-0.02$     &  $1.23-1.36$ \\
    Low-latency flash & $0.15-0.2$  &  $0.02-0.19$  &  $1.19-1.50$  \\
  \bottomrule
  \end{tabular}
\end{table}

\begin{figure}[t]
  \centering
  \includegraphics[height=34mm, trim=0 5 0 0, clip]{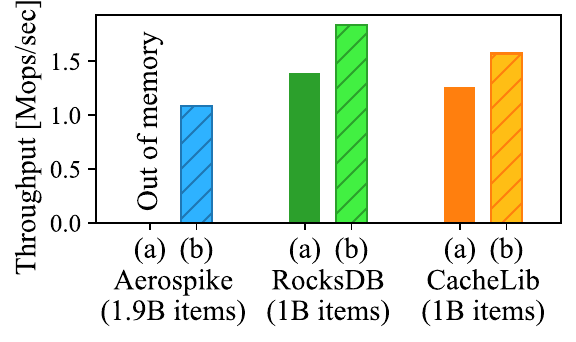}
  \includegraphics[height=34mm, trim=0 377 770 0, clip]{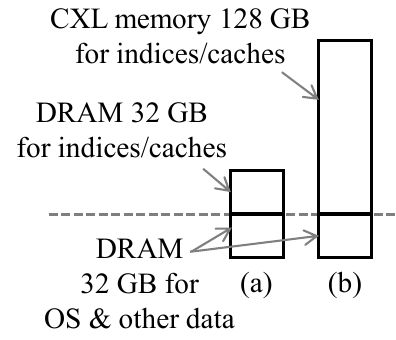}
  \caption{Throughput comparison between cases where space-consuming in-memory data is stored in (a) 32 GB of DRAM and in (b) 128 GB of microsecond-latency memory.}
  \label{fig:extra_memory}
\end{figure}

Instead of reducing the server cost, the saved money could be used to purchase more secondary memory.
We consider another hypothetical scenario where our server had only 32 GB of DRAM.
We could add 128 GB of flash-based CXL memory at a potentially cheaper price than adding another 32 GB of DRAM.
As shown in \myfig{\ref{fig:extra_memory}}, we keep the original 32 GB DRAM for OS and small in-memory data of an SSD-based KV store, and compare the cases where large indices and caches are stored in (a) the extra 32 GB of DRAM or (b) 128 GB of CXL memory with a latency of 5 \usec{} along with the tail latency described above. 
As shown in the figure, Aerospike can handle up to 1.9 billion items with the increased secondary memory capacity, but it runs out of memory with the DRAM-only system.
RocksDB running on 32 cores for a less skewed key distribution (a Zipf exponent of 0.7) sees an improved throughput by 32\% by having a larger block cache.
CacheLib encountering 1 billion items here uses a larger tier-2 (SSD) cache of 512 GB, but it also requires a correspondingly larger tier-1 cache to maintain throughputs, and the system with the secondary memory is 25\% faster than the DRAM-only system.

\subsection{Future Directions}
\label{sec:future}

Our analysis and evaluation stand on a number of assumptions, some of which call for further study and effort as discussed below.

\subsubsection{Software Modification}
This work has modified existing SSD-based KV stores, and reducing the modification labor is out of the scope of this paper.
We believe the extent of the modification has been modest relative to the existing large code bases we have built on, but nonetheless it is desirable to automate the conversion.
We have made all the code available, and we hope our findings will encourage communities to develop SSD-based KV stores that natively support longer-latency memory without requiring modification.

\subsubsection{Analysis Scope and Model Applicability}
Our analysis covers SSD-based KV stores that follow our operation model.
While we have shown that our throughput equation holds for three KV stores that are quite different in design, it is likely that there are KV stores and workloads that significantly deviate from our model.
It would be valuable to further extend our model, and also to take into account more specifics of prefetching \cite{UnderHoodPrefetch} and of interconnects such as CXL \cite{HitchhikerCXL}.
In the meantime, since our operation model only requires there to be memory accesses and IOs, the model may also be applicable to those other than SSD-based KV stores including relational databases and file systems.
Expanding the scope of our analysis and applications is one of the interesting future directions.

\subsubsection{Tiered Memory for Indices and Caches}
This work has studied the scenario where the entire indices and caches are offloaded to secondary memory.
While we have demonstrated near-DRAM throughput in many cases, it would be useful to consider using both of the host DRAM and secondary memory for indices and caches to further mitigate performance degradation.
Designing tiering and migration techniques specifically for microsecond-latency memory is another promising avenue for future work.

\subsubsection{Commercial Availability}
For lack of commercial memory devices having microsecond-level latency at the time of this work, we have used FPGA-based memory, and the commercial viability of microsecond-latency memory remains to be seen.
However, devices equipped with slower memory media are emerging \cite{CMMH}, and we hope our work encourages this trend by demonstrating their potential usefulness based on a research prototype.
Even if the current trend does not lead to near-term commercialization, we believe our work can contribute to the informed decision the industry will be making.
We also hope that, since memory hierarchy is a recurring theme in computer science, our work will add to that knowledge base for when hardware comes back to this trend in the near future again.

This work has shown that SSD-based KV stores are less susceptible to the slowdown coming from the limited prefetch queue depth $P$ in hiding microsecond-level memory latency.
The reason why $P$ is small may be partly because current CPUs expect sub-microsecond memory latency only.
If microsecond-latency memory becomes a commodity, future CPUs may support deeper queues, which would further alleviate the slowdown.

Our FPGA-based memory has been designed to introduce a user-specified fixed latency.
Commercial microsecond-latency memory devices, if realized, would be different in that they would each have a unique tail latency profile depending on their memory media and its controller, and that they would likely implement an on-device cache such that latency would be much shorter upon cache hits.
It would be valuable to study the impact of those features.

\section{Related Work}

The performance impacts of memory expansion and pooling are being actively studied \cite{DemystifyCXL,Pond,SMTSoftMemTiering,TPPTransparentPagePlacement,Elastic,EvalCXLHPC}.
They typically deal with a memory latency of a few hundreds of nanoseconds, and tiering the host DRAM and secondary memory is often effective in curbing performance degradation \cite{SMTSoftMemTiering,TMCTieredMemConfig,TPPTransparentPagePlacement,MemTieringWarehouse,MTMRethink}.
However, some applications are latency-sensitive and suffer non-negligible performance degradation \cite{Pond,EvalCXLHPC}, and mitigating them in general settings still requires a fair amount of host DRAM usage \cite{MemTieringWarehouse}.

By focusing on KV stores, a number of hybrid KV store designs utilizing secondary memory have been proposed \cite{HiKv,CloseGapVolatilePersistent,ExploreSCMKV,FPTree,ChameleonDB,FlashKVSSCM,MatrixKV,SLMDB,Viper,NoveLSM,TriangleKV}, which again assume sub-microsecond latency as they use Optane DCPMM.
Notably, a few of them also use SSDs in addition to secondary memory \cite{NoveLSM,MatrixKV,TriangleKV,FlashKVSSCM}, thus studying the same memory-storage configuration as our work.
Their goals are to improve one type of SSD-based KV store by proposing new techniques that take advantage of sub-microsecond-latency memory devices.
Aside from using secondary memory, some KV stores aim to hide even shorter, host DRAM latency to speed up execution on the host DRAM \cite{CoroBase,MosaicDB}.
Other works explore different storage configurations including two-tier storage \cite{HyperDB,LogStore} and KV-oriented storage devices and use \cite{Dotori,NVMKV}.

There are a few prior works that study much longer, microsecond-level memory latency outside of the application domain of SSD-based KV stores \cite{KillerMicroseconds,VLDBSuzuki,XLGPU}.

Our work is complementary to these prior works in that it studies the performance impact of microsecond-latency memory on varying types of SSD-based KV stores, with the goal of showing they can be made tolerant to microsecond-level memory latency with existing techniques through new analysis and evaluations.

\section{Conclusion}

This paper has studied the impact of using microsecond-latency memory for in-memory indices and caches in SSD-based KV stores.
Our analysis has revealed that the presence of IO significantly enhances latency-tolerance of SSD-based KV stores, and our evaluation has demonstrated that they can achieve near-DRAM throughputs if they employ standard latency-hiding techniques.
This indicates that SSD-based KV stores involving latency-sensitive in-memory traversal can use microsecond-latency memory as a low-cost alternative to DRAM.
We believe these findings provide novel insights into the performance of SSD-based KV stores, and hope that they will encourage communities to expand the application scope of microsecond-latency memory.

\begin{acks}
We would like to thank our anonymous reviewers, as well as Hidekazu Tadokoro, Tetsuya Sunata, and Yohei Hasegawa for their constructive criticism and insightful feedback.
\end{acks}

\bibliographystyle{ACM-Reference-Format}
\bibliography{xlkv}


\end{document}